\documentclass[preprint]{imsart}

\usepackage[latin1]{inputenc}
\usepackage[english]{babel}
\usepackage{enumerate,rotate}
\usepackage{amssymb,latexsym,amsmath,amsthm,mathrsfs,bm}
\usepackage{graphicx,epsfig}
\usepackage{color}
\usepackage{geometry,a4wide}
\usepackage{natbib}
\usepackage{dsfont}
\usepackage{verbatim}

\usepackage{array}
\usepackage{pdfsync}
\usepackage{booktabs}
\usepackage{subfigure}
\usepackage{amsfonts}

\arxiv{}

\startlocaldefs
\newtheorem{definition}{Definition}[section]

\newtheorem{proposition}[definition]{Proposition}

\def\C { \mathcal{C}}

\def�{\mathcal{S}}

\def\a {\alpha}
\def\t {\tau}

\def \th {\theta}

\def\r {\rho}
\def\g {\gamma}

\def �{\Theta}
\def\a {\alpha}
\def\g {\gamma}

\def\t {\tau}

\def�{\Theta}

\def\E {\mathbb{E}}

\def \ba {\boldsymbol{\alpha}}
\def \bp {\mathbf{p}}
\def \bth {\boldsymbol{\theta}}
\def \bg {\boldsymbol{\gamma }}
\def \bj  {\mathbf{j}}
\def \bx {\mathbf{x}}
\def \by {\mathbf{y}}
\def \bt {\mathbf{t}}
\def \bq {\mathbf{q}}

\renewcommand{\Re}{\mathbb{R}}
\newcommand{\indic}{\mathds{1}}
\endlocaldefs

\begin{document}

\begin{frontmatter}

% "Title of the Paper"
\title{Bayesian estimation for a parametric Markov Renewal model applied to seismic data}
%\thankstext{t1}{This is an original survey paper}
\runtitle{Bayesian Markov Renewal model for seismic data}

% indicate corresponding author with \corref{}
% \author{\fnms{John} \snm{Smith}\thanksref{t2}\corref{}\ead[label=e1]{smith@foo.com}\ead[label=e2,url]{www.foo.com}}
% \thankstext{t2}{Thanks to somebody} 
% \address{line 1\\ line 2\\ \printead{e1}\\ \printead{e2}}

\author{\fnms{Ilenia} \snm{Epifani}\corref{}\ead[label=e1]{ilenia.epifani@polimi.it}}
\address{Politecnico di Milano, Piazza Leonardo da Vinci 32, 20133 Milano, Italy\\ \printead{e1}}
\and
\author{\fnms{Lucia} \snm{Ladelli}\ead[label=e2]{lucia.ladelli@polimi.it}}
\address{Politecnico di Milano, Piazza Leonardo da Vinci 32, 20133 Milano, Italy\\ \printead{e2}}
\and
\author{\fnms{Antonio} \snm{Pievatolo}\ead[label=e3]{antonio.pievatolo@cnr.it}}
\address{IMATI-CNR, Via Bassini 15, 20133 Milano, Italy\\ \printead{e3}}
%\and
%\author{\fnms{???} \snm{???}\ead[label=e2]{???}}
%\address{\printead{e2}}

\runauthor{Epifani, Ladelli, Pievatolo}

\begin{abstract}
This paper presents a complete methodology for Bayesian inference on a semi-Markov process, from the elicitation of the prior distribution, to the computation of posterior summaries, including a guidance for its JAGS implementation. The holding  times  (conditional on the transition between two given states)  are assumed to be Weibull-distributed.  We  examine the elicitation of the joint prior density of the shape and scale parameters of the Weibull distributions, deriving  a specific class of priors in a natural way, along with a method for the determination of hyperparameters based on ``learning data'' and moment existence conditions. This framework is applied to  data of earthquakes of three types of severity (low, medium and high size) that occurred in the central Northern Apennines in Italy and collected by the \cite{CPTI04} catalogue. Assumptions on two types of energy accumulation and release mechanisms are evaluated.
\end{abstract}



% history:
% \received{\smonth{1} \syear{0000}}

%\tableofcontents

\end{frontmatter}
\section{Introduction}
\label{sec:introduction}
Markov Renewal processes or their semi-Markov representation have been considered in the seismological literature as models which allow the  distribution of  the inter-occurrence times between earthquakes to  depend on  the last and the next earthquake and to be  not necessarily exponential. 
The time predictable and the slip predictable models studied in \cite{ShimazakiNakata},  \cite{GuagMol},  \cite{GuagMolMul} and  \cite{BetroGaravGuagRotonTagl} are special cases of Markov Renewal processes. These models are capable of interpreting the predictable behavior of strong earthquakes in some seismogenic areas. In these processes the magnitude is a deterministic function of the inter-occurrence time. 
A stationary Markov Renewal process with Weibull inter-occurrence times has been studied   from a classical statistical point of view in \cite{Alvarez}. The Weibull model allows for the consideration of monotonic hazard rates; it contains the exponential model as a special case which gives a Markov Poisson point process. In \cite{Alvarez} the model parameters were fitted to the large earthquakes in the North Anatolian Fault Zone   through maximum likelihood and the Markov Poisson point process assumption was tested.  
In order to capture a non monotonic behavior  in the hazard, in \cite{GaravagliaPavani} the model of Alvarez was modified and a Markov Renewal process with inter-occurrence times that are mixtures of an exponential and a Weibull distribution was fitted to the same Turkish data. 
In \cite{Masala} a parametric semi-Markov model with a generalized Weibull distribution for the inter-occurrence times was adapted to Italian earthquakes. Actually the  semi-Markov model with  generalized Weibull distributed times was first used in \cite{Foucher} to study the evolution of HIV infected patients. \cite{Votsi} considered a semi-Markov model for the seismic hazard assessment in the Northern Aegean sea and  estimated the quantities of interest (semi-Markov kernel, Markov Renewal functions, etc.) through a nonparametric method.

While a wide literature concerning classical inference for Markov Renewal models for earthquake forecasting exists, to our knowledge a Bayesian approach is limited in this context. \cite{PatwardhanEtAlii} considered a semi-Markov model with log-normal distributed discrete inter-occurrence times and applied it to the large earthquakes  in the circum-Pacific belt. 
They stressed the fact that it is relevant to use Bayesian techniques when prior knowledge is available and it is fruitful even if the sample size is small. 
\cite{MarinPlaRiosInsua}  also employed semi-Markov models in the Bayesian framework, applied to a completely different area: sow farm management. They used WinBugs to perform computations (but without giving details) and they elicited their prior distributions on parameters from knowledge on farming practices.

From a probabilistic viewpoint, a Bayesian statistical treatment of a semi-Markov process amounts to  model the data  as  a mixture of semi-Markov processes, where the mixing measure is supported on the parameters, by means of their prior laws. A complete characterization of such a mixture has been given in
\cite{EpifaniFortiniLadelli}.

In this paper we develop a parametric Bayesian analysis for a Markov Renewal process modelling earthquakes in an Italian seismic region. The magnitudes are classified into three categories according to their severity: low, medium and high size, and these categories represent the states visited by the process. As in \cite{Alvarez}, the inter-occurrence times are assumed to be Weibull random variables.  The ``current sample'' is formed by the sequences of earthquakes in a homogeneous seismic region and by the corresponding inter-occurrence times collected up to a time $T$. When $T$ does not coincide with an earthquake, the last observed inter-occurrence time is censored. 
The prior distribution of the parameters of the model is elicited using a ``learning dataset'', i.e. data coming from a seismic region similar to that under analysis.
The posterior distribution of the parameters is obtained through Gibbs sampling and the following summaries are estimated: transition probabilities, shape and scale parameters of the Weibull holding times for each transition and the so-called cross state-probabilities (CSPs). The transition probabilities indicate whether the strength of the next earthquake is in some way dependent on the strength of the last one; the shape parameters of the holding times indicate whether the hazard rate between two  earthquakes of given magnitude classes is decreasing or increasing; the CSPs give the probability that the next earthquake  occurs at or before a given time and is of a given magnitude, conditionally on the time elapsed since the last earthquake and on its magnitude.

The paper is organized as follows. In Section~\ref{sec:1} we illustrate the dataset and  we discuss the choice of the Weibull model in detail. Section~\ref{sec:2} introduces the parametric Markov Renewal model. %Three magnitude classes are considered and a Weibull distribution is assumed for the inter-occurrence times.  
Section~\ref{sec:3} deals with the elicitation of the prior. %, which is based on a generalized inverse Gamma distribution, considering also the case of scarce prior information. 
Section~\ref{sec:apennines} contains the Bayesian data analysis with the estimation of the above-mentioned summaries. We  also test a time predictable and a slip predictable model against the data. Section~\ref{sec:6} is devoted to some concluding remarks. Appendix \ref{sec:gibbsampling} contains the detailed derivation of the full conditional distributions and the JAGS (Just Another Gibbs Sampler) implementation of the Gibbs sampler (\cite{Plummer}).

\section{A test dataset}
\label{sec:1}
\begin{figure}
\includegraphics*[width=10cm]{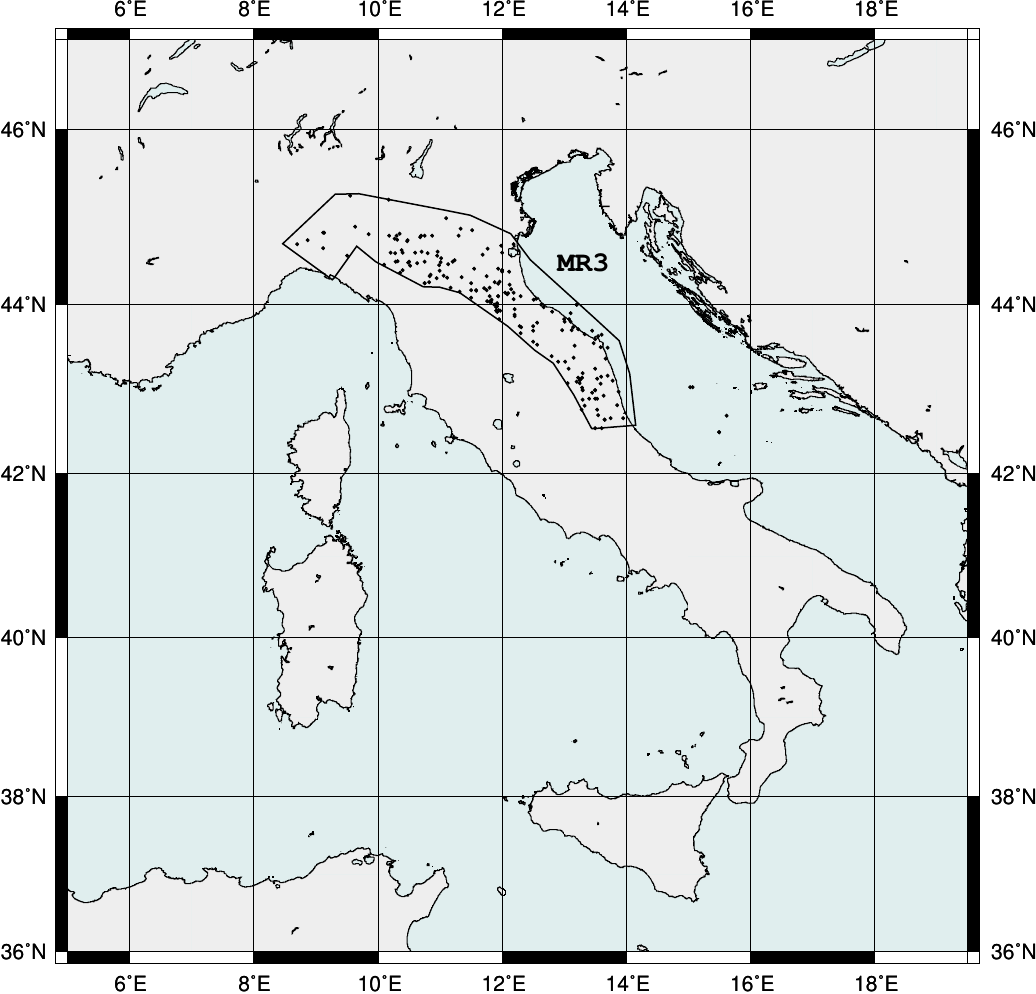}
\caption{Map of Italy with dots indicating earthquakes with magnitude $M_w\ge 4.5$ belonging to macroregion $\text{MR}_3$ (\cite{Rotondi}). Inclusion in the macroregion was based on the association between events and seismogenic sources; the region contour has only an aesthetic function.}
\label{fig:mappa}
\end{figure}

We tested our method on a sequence of seismic events chosen among those examined in \cite{Rotondi}, which was given us by the author. The sequence collects events that occurred in a tectonically homogeneous macroregion, identified as $\text{MR}_3$ by Rotondi and corresponding to the central Northern Apennines in Italy. The subdivision of Italy into eight (tectonically homogeneous) seismic macroregions can be found in the \cite{DISS} and the data are collected in the \cite{CPTI04} catalogue. If one considers earthquakes with magnitude\footnote{We refer to the moment magnitude which is related to the seismic moment $M_0$ by the following relationship: $M_w = \frac{2}{3}(\log_{10}M_0-16.05$); see \cite{Hanks}, where it is denoted by \textbf{M}.} $M_w\ge 4.5$, the sequence is complete from year 1838: a lower magnitude would make the completeness of the series questionable, especially in its earlier part. The map of these earthquakes marked by dots appears in Figure~\ref{fig:mappa}.
As a lower threshold for the class of strong earhquakes, we choose $M_w\ge 5.3$, as suggested by \cite{Rotondi}. Then a magnitude state space with three states is obtained by indexing an earthquake by 1, 2 or 3 if its magnitude belongs to intervals $[4.5,4.9)$, $[4.9,5.3)$, $[5.3,+\infty)$, respectively. Magnitude 4.9 is just the midpoint between 4.5 and 5.3 and the released energy increases geometrically as one moves through the endpoints, with a common ratio of 4: if $M_0(M_w)$ denotes the seismic moment $M_0$ associated with $M_w$, then $M_0(5.3)/M_0(4.9) = M_0(4.9)/M_0(4.5) = 10^{\frac{3}{2} 0.4} \simeq 4$.

The energy released from an earthquake with $M_w=4.9$ does not match the midpoint between seismic moments associated with magnitudes $4.5$ and $5.3$ (in fact, this correspondence holds if $M_w=5.1$). However, there seem to be no general rule in the literature for splitting magnitude intervals. For example, \cite{Votsi} used cut-points 5.5, 5.7 and 6.1, so that $M_0(6.1)/M_0(5.7)\simeq 4$ and $M_0(5.7)/M_0(5.5)\simeq 2$, while the energy midpoint is at $M_w=5.9$; following \cite{Altinok}, \cite{Alvarez} uses  cut-points 5.5, 6.0 and 6.5; \cite{Masala} employed the magnitude classes $M_w<4.7$, $M_w\in [4.7, 5)$, $M_w\ge 5$. All these authors do not give any special reason for their choices.

A more structured approach is attempted by \cite{Sadeghian}, who applied a statistical clustering algorithm to magnitudes, and again by \cite{Votsi} when they propose a different classification of states that combines both magnitude and fault orientation information. From a modelling viewpoint, this latter approach is certainly preferable, because it is likely to produce more homogeneous classes, however we do not have enough additional information to attempt this type of classification of our data in a meaningful way. An entirely different approach is that based on risk, in which cut-points would change with the built environment. 
%As a final remark on this issue, we note that the \cite{CPTI11} seismic map included in the presentation of the catalogue employs a magnitude classification in steps of 0.5, with cut-points starting at 4.25 and reaching 6.75.

\begin{figure}
\includegraphics*[width=10cm]{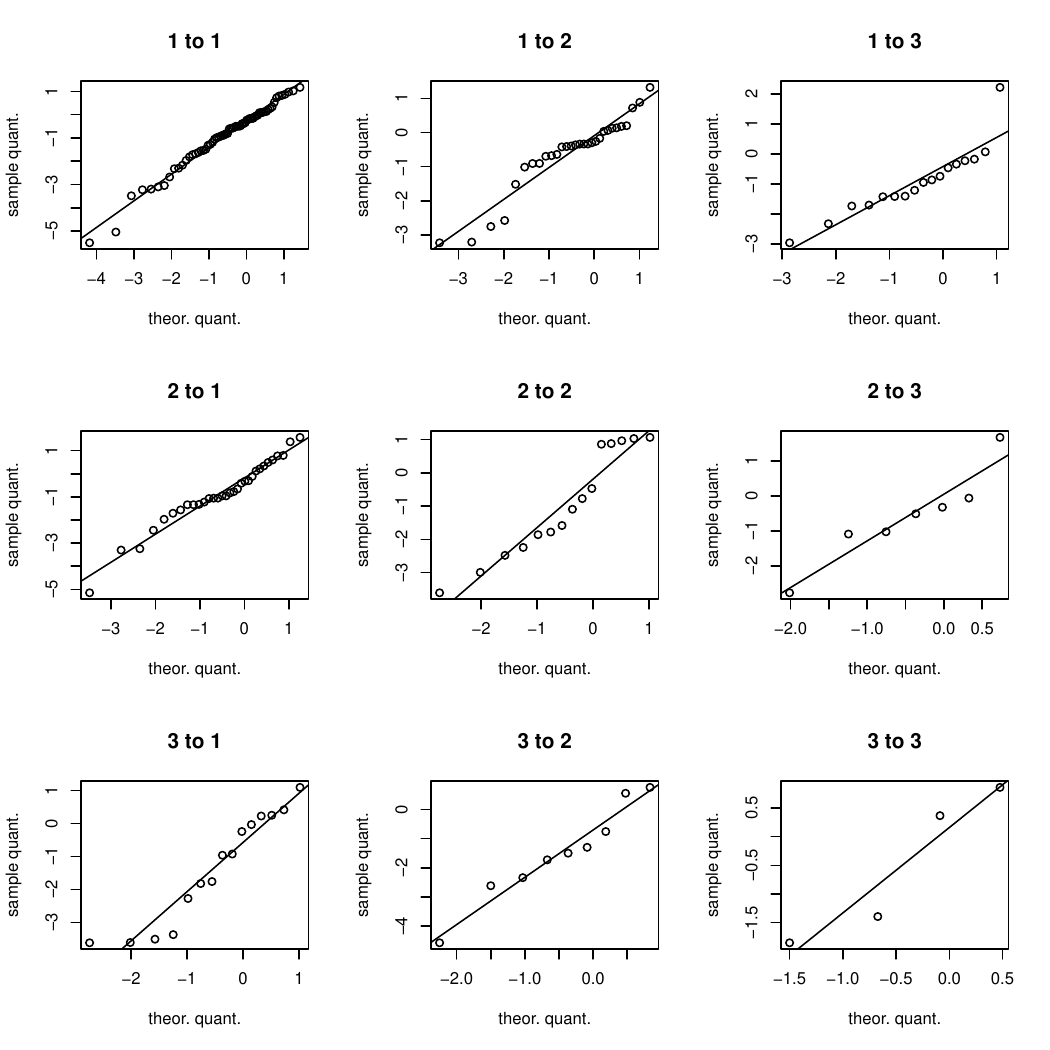}
\caption{Weibull qq-plots of earthquake inter-occurrence times (central Northern Appennines) classified by transition between magnitude classes.}
\label{fig:qqplots}
\end{figure}

We now examine inter-occurrence times. \cite{Rotondi} considers a nonparametric Bayesian model for the  inter-occurrence times between strong earthquakes (i.e.  $M_w\ge 5.3$), after a preliminary data analysis which rules out  Weibull,  Gamma, log-normal distributions among others frequently used. % holding time distributions. 
On the other hand, with a Markov Renewal model, the sequence of all the inter-occurrence times is subdivided into shorter ones according to  the  magnitudes, so that we think that a  parametric distribution is  a viable option. In particular, 
we focussed on the macroregion $\text{MR}_3$ because  the Weibull distribution seems to fit the inter-occurrence times better than in other macroregions. This fact is based on qq-plots. The qq-plots for $\text{MR}_3$ are shown in Figure~\ref{fig:qqplots}. The plot for transitions from 1 to 3 shows a sample quantile that is considerably larger than expected. The outlying point corresponds to a long inter-occurrence time of about 9 years, between 1987 and 1996, while 99~percent of the  inter-occurence times are below 5 years. Obviously, the classification into macroregions influences the way the  earthquake sequence is subdivided.

Given the Markov Renewal model framework, holding time distributions other than the Weibull could be used, such as the inverse Gaussian, the log-normal and the Gamma. However, the inverse Gaussian qq-plots clearly indicate that this distribution does not fit the data. As for the log-normal, the outlying point in the qq-plot of the $(1,3)$ transition becomes only a little less isolated, but at the expense of introducing an evident curvature in the qq-plot of the $(1,1)$ transition, whereas the remaining qq-plots are unchanged. The Gamma qq-plots are indistinguishable from the Weibull qq-plots, but we prefer working with the Weibull in view of the existing literature on seismic data analysis where the Weibull is employed. In this respect, we could follow \cite{Masala} and choose the generalized Weibull, which includes the Weibull, but the qq-plots are unchanged even with the extra parameter. 
From a Bayesian computational point of view,  there is no special reason for preferring the (possibly generalized) Weibull to the Gamma, as neither of them possesses a conjugate prior distribution and numerical methods are needed in both cases for making inference.

In the existing literature, the Weibull distribution has been widely used to model holding times between earthquakes from different areas and with different motivations. In Section \ref{sec:introduction} we mentioned \cite{Alvarez}, \cite{GaravagliaPavani} and \cite{Masala}, but there are also other authors. \cite{AbaimovEtAlii} argued that the  increase in stress caused by the motion of tectonic plates at plate boundary faults is adequately described by an increasing hazard function, such as the Weibull can have. Instead, other distributions have an inappropriate tail behaviour: the log-normal hazard tends to zero with time and the inverse Gaussian hazard tends to a constant. Goodness-of-fit checks for the recurrence times of slip events in the creeping section of the San Andreas fault in central California confirmed that the Weibull is preferable to the mentioned alternatives. \cite{HrostMousl} considered a Weibull model, for single faults (or fault systems with homogeneous strength statistics) and power law stress accumulation. They derived the Weibull model from a theoretical framework based on the statistical mechanics of brittle fracture and they applied it to microearthquake sequences (small magnitudes) from the island of Crete and  from a seismic area of Southern California, finding agreement with the data except for some deviations in the upper tail. Regarding tail behaviour, we can make a connection with
\cite{HasumiEtAlii}, who analyzed a catalogue of the Japan Meteorological Agency. These data support the hypothesis that the holding times can be described by a mixture of a Weibull distribution and a log-Weibull distribution (which possesses a heavier tail); if only earthquakes with a magnitude exceeding a threshold are considered, the weight of the log-Weibull component becomes negligible as the threshold increases.

\section{Markov Renewal  model}
\label{sec:2}
Let us observe, over a period of time  $[0,T]$, a process in which different events occur, with random inter-occurence times.
Let us suppose that the possible states of the process are the points of a finite set 
$E=\{1,\dots , s \}$ and that  the process starts from state $j_0$.
Let us denote by $\t $ the number of times the process changes states in the time interval $[0,T]$ and by
$t_i$  the time of the $i$-th change of state. Hence,
$0<  t_1< \dots < t_{\t }\leq T$.  Let  $j_0, j_1,\dots, j_{\t }$ be the sequence of states visited by
the process and $x_i$ the holding time in the state $j_{i-1}$, for $i=1,\dots, \t $. Then
\begin{equation*}
%  \label{eq:intertempi}
x_i= t_i-t_{i-1} \qquad \mbox{for} \;  i=1,\dots,  \t 
\end{equation*}
with $t_0:=0$.  Furthermore, let $u_T$ be  the time spent in $j_\tau$
\begin{equation*}
%  \label{eq:tempo_cens}
u_T= T-t_{\t } , 
\end{equation*}
so the time $u_T$ is a right-censored time. % if $\t(h)<T$. 
Finally,  our data are collected in the vector $(\bj, \bx, u_T)$, where $	(\bj, \bx) =(j_{n}, x_{n})_{n=1,\ldots, \t   }$.

In what follows, we assume that the data $(\bj, \bx, u_T)$ are the result of the observation %, over the time interval $[0,T]$,
of a homogeneous Markov Renewal process $(J_n, X_n)_{n\geq 0}$ starting from $j_0$.
This means that the sequence $(J_n, X_n)_{n\geq 0}$ satisfies
\begin{equation}
		P(J_0=j_0)  = 1, \quad  P(X_0=0)  =1 
 \label{eq:MRP_J_0}
\end{equation}
and for every $n\geq 0$, $j\in E$ and $t\geq 0$
\begin{equation}
  \label{eq:MRP_n}
P(J_{n+1}=j, X_{n+1}\leq t | (J_{k}, X_{k})_{k\leq n})=
P(J_{n+1}=j, X_{n+1}\leq t | (J_{n}, X_{n}))= p_{J_{n} j} F_{J_{n}  j}(t) \ .
\end{equation}
The transitions probabilities $p_{ij}$'s are collected in a transition matrix $\bp = (p_{ij})_{i,j \in E}$  and $(F_{ij})_{i,j \in E}$ is an array of
distribution functions on $\Re_{+}=(0,+\infty)$.
For more details on Markov Renewal processes  see, for example, \cite{LimniosOprisan}. 
We just recall that, under  Assumptions~\eqref{eq:MRP_J_0} and \eqref{eq:MRP_n}:
\begin{itemize}
	\item[--] the process $(J_n)_{n\geq 0}$ is a Markov chain, starting from $j_0$,  with transition matrix $\bp$,
	\item[--] the holding  times $(X_n)_{n\geq 0}$, conditionally on $(J_n)_{n\geq 0}$, form a sequence of independent positive random variables, with distribution function $F_{J_{n-1}\ J_n}$. 
\end{itemize}
We assume that the functions $F_{ij}$ are absolutely continuous with respect to the Lebesgue measure 
with density $f_{ij}$. Hence, the likelihood function of the data $(\bj, \bx, u_T)$ is
\begin{equation}
\label{eq:likelihood_general}
L(\bj, \bx, u_T)=
\left(\prod_{i=0}^{\t-1}p_{j_i j_{i+1}}f_{j_i j_{i+1}}(x_{i+1})\right)^{\indic(\t >0)}\times 
\sum_{k\in E}p_{j_{\t}k} \bar{F}_{j_{\t}k}(u_T),
\end{equation}
where, for every $x$, $\bar{F}_{ij}$ is the survival function 
\begin{displaymath}
\bar{F}_{ij}(x)=1-F_{ij}(x)=P(X_{n+1}>x|J_n=i, J_{n+1}=j) \ .
\end{displaymath}
Furthermore, we assume that each inter-occurrence time has a  Weibull density $f_{ij}$  with
shape parameter $\a_{ij}$ and scale parameter $\theta_{ij}$, i.e. 
\begin{equation}
  \label{eq:weib}
f_{ij}(x)=\frac{\a_{ij}}{\theta_{ij}}\left(\frac{x}{\theta_{ij}}\right)^{\a_{ij}-1}
\exp\left\{-\left(\frac{x}{\theta_{ij}}\right)^{\a_{ij}}\right \} \ , \quad x>0 , \; \a_{ij}>0, \;\theta_{ij}>0 \ .
\end{equation}
For conciseness, let 
	${\boldsymbol{\alpha}}=(\a_{ij})_{i,j\in E}$ and ${\boldsymbol{\theta}}=(\th_{ij})_{i,j\in E}$.

In order to write  the  likelihood in a more convenient way, let us introduce the following natural statistics. We will say that the process visits the string $(i,j)$ if a visit to  $i$ is followed by a visit to  $j$ and we denote by
\begin{itemize}
	\item[--] $x^{\r}_{ij}$ the time spent in  state $i$ at the $\r$-th visit to the string $(i,j)$,
	\item[--] $N_{ij}$ the number of visits to the string $(i,j)$.
\end{itemize}
Then, assuming $\t \ge 1$, Equations~\eqref{eq:likelihood_general} and \eqref{eq:weib} yield the following 
representation of  the likelihood function 
\begin{multline}
\label{eq:likelihood-param}
%L(\bj, \bx)=  
L(\bj, \bx, u_T \, |\, \bp, \boldsymbol{\a},\boldsymbol{ \theta}) =\prod_{i,k\in E}p_{ik}^{N_{ik}}\times \\
\times  \prod_{i,k\in E}\left[\a_{ik}^{N_{ik}}\frac{1}{\theta_{ik}^{\a_{ik}N_{ik}}}\left(\prod_{\r =1}^{N_{ik}}x^{\r}_{ik}\right)^{\a_{ik}-1} \times 
\exp\left\{-\frac{1}{\theta_{ik}^{\a_{ik}}}\sum_{\r  =1}^{N_{ik}}(x^{\r}_{ik})^{\a_{ik}}\right\}\right]\times \\
 \times \left(\sum_{k\in E}p_{j_{\t}k}
\exp\left\{-\left(\frac{u_T}{\theta_{j_{\t} k}}\right)^{\a_{j_{\t}  k}}\right\}\right)\ .
\end{multline}
Our purpose is now to perform a Bayesian analysis for $\bp, \boldsymbol{\alpha}$ and $ \boldsymbol{\theta}$ which allows us to introduce prior knowledge on the parameters. As shown in Appendix \ref{sec:gibbsampling}, this analysis is possible via a Gibbs sampling approach.

\section{Bayesian analysis}
\label{sec:3}

\subsection{The prior distribution}
\label{subsec:3.1}
Let us assume that a priori  $\bp$ is independent of $\ba$ and $\bth$. In particular, 
 the rows of $\bp$ are $s$ independent vectors with Dirichlet distribution
with parameters $\bg_1, \cdots ,\bg_s$ and total mass $c_1, \cdots,c_s$, respectively. This means
that, for $i=1,\ldots ,s$, the prior density of the $i$-th row is
\begin{equation}
	\label{eq:dirich}
	\pi_{1,i}(p_{i1},\ldots , p_{is}) = \frac{\Gamma(c_i)}{\prod_{j=1}^{s}\Gamma(\g_{ij})} {\prod_{j=1}^{s}p_{ij}^{\g_{ij}-1}}
\end{equation}
on $T=\{ (p_{i1},\ldots , p_{is})| \ p_{ij} \geq 0,\  \sum_j p_{ij} =1\}$  where $\bg_i =(\g_{i1}, \cdots , \g_{is})$, with  $\g_{ij}>0$ and $c_i=\sum_{j=1}^s \g_{ij}$. 

As far as $ \ba$ and $\bth$ are concerned,  the $\th_{ij}$'s, given the $\a_{ij}$'s, are independent with generalized inverse Gamma
densities
\begin{equation}
	\label{eq:pi2}
	\pi_{2,ij}(\th_{ij}| \ba) = \pi_{2,ij}(\th_{ij}| \a_{ij}) = 
	\frac{\a_{ij}b_{ij}(\a_{ij})^{m_{ij}}}{\Gamma(m_{ij})}\th_{ij}^{-(1+m_{ij}\a_{ij})}
	\times \exp\left\{-\frac{b_{ij}(\a_{ij})}{\th_{ij}^{\a_{ij}}}\right\}, 
	\;\;  \th_{ij}>0,
\end{equation}
where  $m_{ij}>0$ and
\begin{equation}
	\label{eq:b-q}
	b_{ij}(\a_{ij}) = \left(t^{ij}_{q_{ij}}\right)^{\a_{ij}}[(1-q_{ij})^{-1/m_{ij}}-1]^{-1}
\end{equation} 
with 
$t^{ij}_{q_{ij}}>0$ and $q_{ij}\in (0,1)$.
In other terms, $\th_{ij}^{- \a_{ij}}$, given $ \a_{ij}$, has a prior Gamma density with shape $m_{ij}$ and  scale $1/b_{ij}(\a_{ij})$. In symbols $\th_{ij}|
\a_{ij} \sim \mathcal{ GIG}(m_{ij}, b_{ij}(\a_{ij}), \a_{ij})$. 
We borrow the expression of the $b_{ij}(\a_{ij})$'s in \eqref{eq:b-q} from \cite{Bousquet2} and, as a consequence of this choice,  $t^{ij}_{q_{ij}}$ turns out to be the  marginal quantile of order $q_{ij}$ of an inter-occurrence time between states $i$ and $j$. Indeed, if $\pi_{3,ij}$ denotes the density of $\a_{ij}$ and $X$ is such a random time, then 
\begin{align*}
	P(X> t) & = \int_{0}^{+\infty}\int_{0}^{+\infty}
					P(X> t|\a_{ij},\theta_{ij})\pi_{2,ij}(\th_{ij}|\a_{ij})\pi_{3,ij}(\a_{ij})d\th_{ij} d\a_{ij}\\
				 & = \int_{0}^{+\infty}\Big[\frac{b_{ij}(\a_{ij})}{b_{ij}(\a_{ij})+t^{\a_{ij}}}\Big]^{m_{ij}}\pi_{3,ij}(\a_{ij})d\a_{ij} \ , 
				\qquad \qquad \qquad \qquad\;\forall t>0.
\end{align*}
Hence, in view of \eqref{eq:b-q}, if $t=t^{ij}_{q_{ij}}$, we obtain $P(X>t^{ij}_{q_{ij}})=1-q_{ij}$, for every proper prior density $\pi_{3,ij}$.

Finally, a priori, the 
components of $ \ba$ are independent and have % a log-concave
densities $\pi_{3,ij}$  %with support bounded away from zero
such that
\begin{multline}
	\label{eq:pi3}
	\pi_{3,ij}(\a_{ij})\propto 
		\a_{ij}^{m_{ij}-c_{ij}}\left(\a_{ij}-\a_{0,ij}\right)^{c_{ij}-1}\exp\{-m_{ij}d_{ij}\a_{ij}\}\indic(\a_{ij}\ge\a_{0,ij}), \\  
		\a_{0,ij}\geq 0,\; c_{ij}> 0,\; m_{ij}>0, \; d_{ij}\ge 0\ .
\end{multline}
As far as the prior $\pi_{3,ij}$ is concerned, it is easy to  see that:
\begin{enumerate}[$a)$]
	\item  if $d_{ij} >0$, then $\pi_{3,ij}$ is a proper prior;
	\item if $\a_{0,ij}=0$ and $d_{ij} >0$, %$, \; m_{ij}>0$ and $d_{ij}>0$, 
			then $\pi_{3,ij}$ is a Gamma density; 
	\item if $c_{ij}=1 ,  \ \a_{0,ij}>0 \mbox{ and } d_{ij} >0$,  %, \; m_{ij}>0$ and $d_{ij}>0$, 
		then $\pi_{3,ij}$ is a Gamma density %constrained to be greater than $\a_{0,ij}$; 
		truncated from below at $\a_{0,ij}$;
	\item if $c_{ij}=m_{ij}, \ \a_{0,ij}>0 \mbox{ and } d_{ij} >0$, % \; m_{ij}>0$ and $d_{ij}>0$, 
		then $\pi_{3,ij}$ is a Gamma density shifted by %the quantity 
	$\a_{0,ij}$;
	\item if $c_{ij}=1$ and $m_{ij}\to 0$, then $\pi_{2,ij}(\theta_{ij}|\a_{ij})\pi_{3,ij}(\a_{ij})$ approaches the Jeffreys prior for the Weibull model: $1/\theta_{ij}\indic{(\theta_{ij}>0)}\indic{(\alpha_{ij} \geq \a_{0,ij})}$;
	\item  if  $c_{ij} \ge 1$ \mbox{ and } $m_{ij}\ge 1$, then  $\pi_{3,ij}$ is a log-concave function. % for every $\a_{0,ij}\geq 0$, $c_{ij} \ge 1$, $m_{ij}\ge 1$ and $d_{ij}>0$.
\end{enumerate}
The prior corresponding to the choices in $c)$ was first introduced in \cite{Bousquet} and \cite{Bousquet2}. 
As discussed in \cite{GilksWild}, the log-concavity of $\pi_{3,ij}$ is necessary
in the implementation of the Gibbs sampler (see also \cite{BergerSun}), although adjustments exist for the non-log-concave case (see \cite{GilksBestTan}). 
Furthermore, we will show later that a support suitably bounded away from zero ensures the existence of the posterior moments of the $\th_{ij}$'s.

\subsection{Elicitation of the hyperparameters}

In this section we focus our attention on the prior of $ (  \a_{ij},\th_{ij})$, for  fixed $i,j$. Adapting the approach developed by Bousquet to our situation, we give a  statistical justification of the prior introduced in Subsection~\ref{subsec:3.1}. An interpretation of the hyperparameters is also provided.

For the sake of semplicity,  let us drop the indices $i,j$  in all the notations and  quantities. % introduced in Section~\ref{subsec:3.1}.

Suppose that a ``learning dataset'' $\by_{m} = (y_{1},\ldots , y_{m})$ of $m$ holding times in the state $i$ followed by a visit to the state $j$ is available from another seismic region similar to the one under analysis. 
Therefore the prior scheme defined by  Equations \eqref{eq:pi2}--\eqref{eq:pi3} can be interpreted as a suitable modification of a posterior distribution of $( \a, \th)$, given the learning dataset $\by_{m}$. %,  \eqref{eq:pi2}, \eqref{eq:b-q} and 
This approach allows us to elicit the hyperparameters.

More precisely, consider 
for $(\a, \theta)$ the posterior density, conditionally on $\by_{m}$, when we start from the following improper prior:
\begin{equation}
	\label{eq:pi_alpha1}
\tilde{\pi}(\a, \theta) \propto \theta^{-1}\left(1-\frac{\alpha_0}{\alpha}\right)^{c-1} \indic{(\theta \geq 0)} \indic{(\alpha \geq \alpha_0)} \ , 
\end{equation}
for some suitable $c\geq 1$ and $\alpha_0\geq 0$ (The condition $c\geq 1$ guarantees that $\tilde{\pi}(\a, \theta)$ is a log-concave function with respect to $\alpha$).
Consequently, the posterior density of $\theta$, given $\alpha$ and $\by_{m}$, is
\begin{equation}
	\label{eq:prior-theta}
	\tilde\pi_2(\theta| \by_{m}, \alpha ) = \mathcal{ GIG}(m, \tilde b(\by_{m} ,\alpha), \alpha)
\end{equation}
and the posterior density of $\alpha$ is
\begin{equation}
	\label{eq:prior-alpha}
	\tilde\pi_3(\alpha |\by_{m}) \propto \frac{\alpha^{m-c}(\alpha -\alpha_0)^{c-1}}{\tilde b^m(\by_{m}, \alpha)}
	\exp\{{-m}{\beta(\by_{m})\alpha}\}\indic{(\alpha \geq \alpha_0)} \ , 
\end{equation}
with $\tilde b( \by_{m},\alpha)=\sum_{i=1}^{m}y^{\alpha}_{i}$ and $\beta(\by_{m})= {\sum_{i=1}^{m}\ln y_{i}}/{m}$.
%\begin{remark}
%	\label{remark:123}

Notice that  the posterior we obtain
%$\tilde\pi_2(\theta| \by_{m}, \alpha ) \tilde\pi_3(\alpha |\by_{m})$ 
has a simple hierarchical structure: $\tilde\pi_2(\theta|\by_{m},\alpha)$ is a generalized inverse Gamma density and  this provides both a justification of the form  of the $\pi_2(\theta|\alpha)$ in \eqref{eq:pi2} and an interpretation of the first parameter $m$. Indeed $m$ is equal to  the size of the learning dataset $\by_m$ and so it is a measure of prior uncertainty.

Now, if  we replace the function $\tilde b(\by_{m},\alpha )$  in \eqref{eq:prior-theta} and \eqref{eq:prior-alpha} by the easier convex function of $\alpha$ introduced in \eqref{eq:b-q}, i.e.
$b(\alpha)= t^{\alpha}_q[(1-q)^{-1/m}-1]^{-1}$, with  $t_q>0$ and $q\in (0,1)$, 
then  $\tilde\pi_3(\alpha |\by_{m})$ takes the same form as %the $\pi_3$
in \eqref{eq:pi3} with 
\begin{equation}
	\label{eq:d_ij}
	d=\ln {t}_q- \frac{\sum _{i=1}^{m}\ln y_{i}}{m} \ .
\end{equation}
In this way, we obtain  a justification  of the form  of the prior densities $\pi_{3,ij}$'s in \eqref{eq:pi3} and an easy way to elicit its parameter $d_{ij}$ when the learning dataset is available.
Furthermore, $b(\alpha)$ can be also elicited once the predictive quantile $t_q$ is specified. Its specification can be accomplished, for example, in the two following different ways:
\begin{enumerate}
\item we estimate an empirical quantile $\hat t_q$ from  the learning dataset;
\item an expert is asked about the chance, quantified by $q$, of an earthquake before $t_q$.
%\item if we know nothing about the interoccurrence times, we can represent in the model the absence of information, choosing 
%\[q=0.5\quad\mbox{and}\quad t_{0.5}\sim U(t_1,t_2)
%\]
%that is $t_{0.5}$ uniformly distributed over a big time interval $(t_1,t_2)$, independent from everything else. See Subsection~\ref{sec:scarce} for more details.
\end{enumerate} 
%See  \cite{Bousquet2}. Hence, for a fixed $q\in (0,1)$,   we can estimate $t_q$ using the learning data. Denote by $\hat{t}_q$ such an estimate
%and let
%\begin{equation*}
%	\label{eq:b_q^*}
%		\hat{b}_q(\alpha,\bx_{-m})=\hat{t}_q^{\alpha}[(1-q)^{-1/m}-1]^{-1}.
%\end{equation*}
%Let us denote by $\hat{t}_q$ the estimation of $t_q$  considered and by $ \hat{b}(\alpha %,\by_{m}
%)$ the resulting $b(\a)$.
\mbox{}\indent In the following, if a learning dataset of size $m\ge 2$ is available, we consider an empirical quantile $\hat{t}_q$ of order $q$ such that
\begin{displaymath}
	\ln \hat{t}_q- \frac{\sum _{i=1}^{m}\ln y_{i}}{m}> 0\ .
\end{displaymath}
Therefore, letting $\hat{b}(\alpha)$ denote the value of $b(\a)$ corresponding to $\hat{t}_q$, we propose a Bayesian analysis based on the prior
\begin{equation}\label{pi-2}
	\pi_2(\theta|\alpha )=\mathcal{ GIG}(m, \hat{b}(\alpha), \a)
\end{equation}
and 
\begin{equation}\label{pi-3}
	\pi_3(\alpha)\propto \a^{m-c}(\a -\a_0)^{c-1} \exp\left\{-m\left(\ln \hat{t}_q- \frac{\sum _{i=1}^{m}\ln y_{i}}{m}\right)\a\right\}\indic{(\a \geq \a_0)} \ ,
\end{equation}
where $m$ is the size of the learning dataset. In addition, we choose $c=m$ so that $\pi_3(\alpha)$ is a shifted Gamma prior and consequently it is proper and log-concave.
%Notice that such a choice is possible for almost all set of data.

The remaining hyperparameter $\a_0$ is chosen so that the posterior second moment of $\theta$ is finite. If $\a$ is bounded away from zero, then
\begin{displaymath}
	\E(\theta^2) = \E\left(\frac{\Gamma(m-2/\a)}{\Gamma(m)}\left[\hat{b}(\alpha)\right]^{2 / \a}\right)\leq \tilde{K} \E(\Gamma(m-2/\a))
\end{displaymath}
for a suitable constant $\tilde{K}$. As a consequence if $\a_0=2/m$, then $\E(\theta^2)<+\infty$ and hence also the posterior second moment of $\theta$ is finite. 

The choice $\a_0=2/m$ is suitable only if $m>2$. 
If $m=2$, then  
%the elicitation of the hyperparameters differs from the one presented in the previous subsection (case $m>2$) only for the choice of $\a_0$. In this case the choice 
$\a_0=2/m=1$ and decreasing hazard rates are ruled out.  %no more suitable since it rules out decreasing hazard rates. 
In the absence of additional specific prior information, this is an arbitrary restriction, so a value for $\a_0$ smaller than 1 must be chosen. Then, the prior second moment of $\theta$ is not finite anymore. On the other hand, for the posterior second moment to be finite, we need $\alpha>2/(2+N)$, where $N$ is the number of transitions between the two concerned states in the (current) sample.  Thus the second moment of $\theta$ can stay non-finite, even a posteriori, if $2/(2+N)>\alpha_0$. This would show that the data add little information for that specific transition. To avoid this, we may let $\a_0$ be the minimum between the value $2/3$, corresponding to the smallest learning sample size such that $\a_0<1$, and  the value $2/(2+N)$, necessary for the finiteness of the posterior second moment.  Therefore, $\displaystyle \alpha_0 = \min\{2/3, 2/(2+N)\}$.

Finally, if $\gamma$ denotes the hyperparameter corresponding to the indexes $i$ and $j$ in the Dirichlet prior \eqref{eq:dirich}, then we select $\gamma=m+1$, i.e. $\gamma$ is equal to the number of transitions from state $i$ to state $j$ in the learning dataset, plus one.

\subsection{Scarce prior information}\label{sec:scarce}
The construction of the prior distribution of $(\alpha,\theta)$ must be modified for those pairs of states between which no more than one transition was observed in the learning dataset.

If $m=1$, the single learning observation $y_{1}$ determines $\hat{b}(\alpha)$. As $\hat{t}_q=y_{1}$ for any $q$, it seems reasonable to use $q=0.5$, so $y_{1}$ would represent the prior opinion on the median holding time.  Since $d=0$  when $m=1$, then  $\pi_3(\alpha)$ is improper for any $c>0$. We make it proper by restricting its support to an interval $(\alpha_0,\alpha_1)$. The value $\alpha_1=10$ is suitable for all practical purposes. As before, the choice $\alpha_0 = 2/m=2$ would be too much restrictive, so we select again  $\displaystyle \alpha_0 = \min\{2/3, 2/(2+N)\}$. With regard to $c$, we put $c=2$.
Furthermore, the elicitation of the hyperparameter of the Dirichlet prior is again $\gamma = m+1=2$, i.e. the number of transitions observed in the learning dataset (just one) plus one.

If $m=0$, the prior information on the number of transitions is that there have been no transitions, but there is no information on the holding  times. 
In this case  we can represent  in the model the absence of information, choosing 
\[
	q=0.5\quad\mbox{and}\quad \tilde t_{0.5}\sim U(t_1,t_2),
\]
that is $\tilde t_{0.5}$ is uniformly distributed over a big time interval $(t_1,t_2)$, independently from everything else. 
%If, for example, some data about the inter-occurrence times between other couples of states are available, then  $\tilde{t}_{0.5}$ will be taken uniformly distributed  between the smallest and the largest inter-arrival times observed in the entire learning dataset. 
Hence, we use $q=0.5$ and $\tilde{t}_{0.5}$ to obtain $\hat{b}(\alpha)$ and we fall in the previous case by substituting $m=1$ to $m=0$. 

For clearness,  in Table~\ref{table:hyperparameters}, we summarize the  hyperparameter selection for priors \eqref{eq:dirich}, \eqref{pi-2} and \eqref{pi-3}.
\begin{table}[h]
\renewcommand*{\arraystretch}{1.4}
 \begin{tabular}{r|cccc}
 & $m>2$ & $m=2$ & $m=1$ & $m=0$\\
\hline
$t_q$ & $\hat{t}_q$ & $\hat{t}_q$ & $y_{1}$ & $\tilde{t}_q\sim U(t_1,t_2)$\\
$c$ & $m$ & $m$ & 2 & 2\\
$\a_0$ & $\frac{2}{m}$ &  $\min\left\{\frac23, \frac2{2+N}\right\}$ & $\min\left\{\frac23, \frac2{2+N}\right\}$ & $\min\left\{\frac23, \frac2{2+N}\right\}$ \\ 
$\gamma$ & $m+1$ & $m+1$ & $m+1$ & $2$ \\
\end{tabular}
\caption{Hyperparameter selection as the learning sample size $m$ varies.}
\label{table:hyperparameters}
\end{table}

\section{Analysis of the central Northern Apennines sequence}\label{sec:apennines}
In this section we analyze the macroregion $\text{MR}_3$ sequence, using the semi-Markov model.

We coded the Gibbs sampling algorithm in the JAGS software package, which is designed to work
closely with the \cite{Rpackage} package, in which all statistical computations and
graphics were performed. Details of the Gibbs sampler are in Appendix \ref{sec:gibbsampling}. On the whole, 750,000 iterations for one chain were run for estimating the unknown
parameters in the model, and the first 250,000 were discarded as burn-in.
After the burn-in, one out of every 100 simulated values was kept for
posterior analysis, for a total sample size of 5,000. The
convergence diagnostics, such as those available in the R package CODA
(Geweke, Heidelberger and Welch stationarity test, interval halfwidth
test), were computed for all parameters, indicating that convergence
has been achieved.

Model fitting, model validation and an attempt at forecasting involve the following steps:
\begin{enumerate}
 \item the learning dataset for the elicitation of the prior distribution is chosen;
 \item model fit is assessed by comparing observed inter-occurrence times (grouped by transition) to posterior predictive intervals;
 \item cross state-probabilities are estimated, as an indication to the most likely magnitude and time to the next event, given information up to the present time;
 \item an interpretation in terms of slip predictable or time predictable model is provided.
\end{enumerate}

For the elicitation of the prior distribution, the learning data are taken from $\text{MR}_4$, another macroregion among those considered by \cite{Rotondi}, who examines statistical summaries of the holding times and suggests that $\text{MR}_4$ could be used as a learning set for the hyperparameters of $\text{MR}_3$. \cite{PeruggiaSantner}, in their analysis of the magnitudes and of the inter-occurence times of eartquakes from another Italian area, chose a subset of the incomplete older part of their series to elicit prior distributions. This procedure is justifiable in their case because the old and the new part of the series can be regarded as two different processes and the cut-point between them appears to be clearly identified. If we did the same with our series, we would alter the Bayesian learning process, because we would obtain different posterior distributions on changing the cut-point position.

Transition frequencies and median inter-occurrence times appear in Table \ref{table:empirical.P} for both the $\text{MR}_3$ and the $\text{MR}_4$ datasets. The Dirichlet hyperparameters $\bg_1, \cdots ,\bg_s$ are equalled to the rows of Table \ref{table:empirical.P}\subref{table:0.empirical.P} plus one. The medians are reported because we have selected $q=0.5$ in Table \ref{table:hyperparameters}: the medians in Table \ref{table:empirical.P}\subref{table:0.median} are smaller than the medians in Table \ref{table:empirical.P}\subref{table:1.median} in six entries out of nine, in some cases considerably. 
\begin{table}[h]
	\begin{center}
\subtable[\label{table:1.empirical.P}]{
	\scalebox{ 1}{	\begin{tabular}{rrrr}
	  \hline
	 & to 1 & to 2 & to 3 \\ 
	  \hline
from	1 & 65 &  30 &  17 \\ 
from	  2 &  32 &  15 &   7 \\ 
from	  3 &  15 &   9 &   4 \\ 
	   \hline
	\end{tabular}
}}\quad
\subtable[\label{table:0.empirical.P}]{
	\scalebox{ 1}{	\begin{tabular}{rrrr}
	  \hline
	 & to 1 & to 2 & to 3 \\ 
	  \hline
	from 1 & 114 &  51 &  13 \\ 
	from 2 &  56 &  25 &  4 \\ 
	from 3 &  8 &   8 &   3 \\ 
	   \hline
	\end{tabular}
}}\quad\\
\subtable[\label{table:1.median}]{
	\scalebox{ 1}{	\begin{tabular}{rrrr}
	  \hline
	 & to 1 & to 2 & to 3 \\ 
	  \hline
from	1 & 204 &  257 &  141 \\ 
from	  2 &  150 &  122 &   219 \\ 
from	  3 &  142 &   82 &   309 \\ 
	   \hline
	\end{tabular}
	
}}\quad
\subtable[\label{table:0.median}]{
	\scalebox{ 1}{	\begin{tabular}{rrrr}
	  \hline
	 & to 1 & to 2 & to 3 \\ 
	  \hline
	from 1 & 105 &  61 &  193 \\ 
	from 2 &  104 &  99 &  76 \\ 
	from 3 &  209 &   117 &   78 \\ 
	   \hline
	\end{tabular}
}}
	\end{center}
	\caption{Summaries of datasets $\text{MR}_3$ (tables on the left) and $\text{MR}_4$ (tables on the right);  $\text{MR}_4$ is the learning dataset used for hyperparameter elicitation. (a) and (b): number of observed transitions; (c) and (d): median inter-occurrence times (in days).}
	\label{table:empirical.P}	
\end{table}

\begin{figure}[ptbh]
\centering
\includegraphics[scale=0.5]
{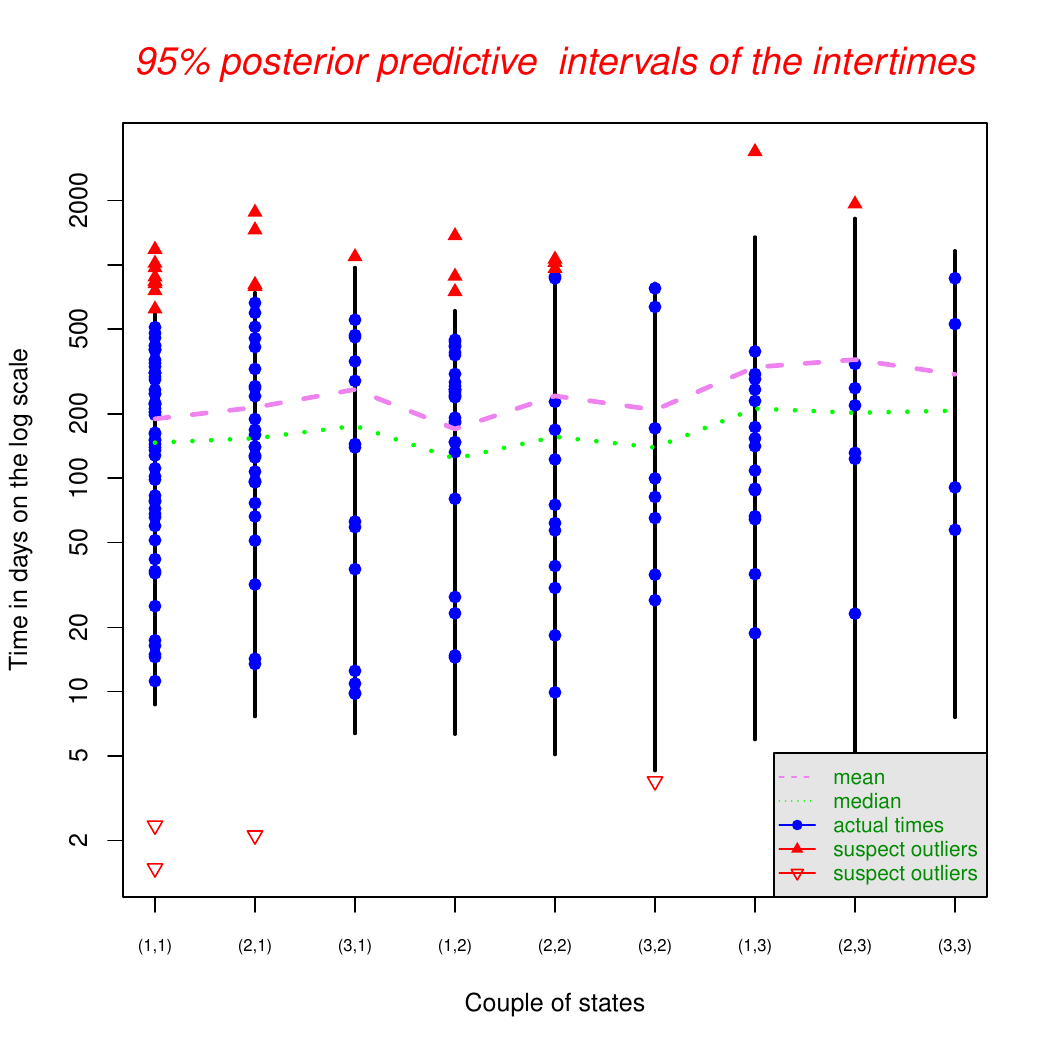}
\caption{Posterior predictive $95$~percent credible intervals of the inter-occurrence times  in days with actual times denoted by (blue) solid dots. Suspect outliers are denoted by (red)-pointing triangles.  The (green) dotted line shows the posterior median and the (violet) dashed line the posterior mean. The prior distribution was elicited from the $\text{MR}_4$ learning set.}
\label{figure:caterpillar_times}
\end{figure}

Let us consider the predictive check mentioned above. Figure~\ref{figure:caterpillar_times} shows posterior predictive 95~percent probability intervals of the inter-occurrence times for every transition, with the observed inter-occurrence times superimposed. These are empirical intervals computed by generating stochastic inter-occurrence times from their relevant distributions at every iteration of the Gibbs sampler. Possible outliers, represented as triangles, are those times with Bayesian $p$-value (that is the predictive tail probability) less than 2.5~percent.
In Table~\ref{table:posterior.time} we report the expected value (and the standard deviation) of the inter-occurrence times. In Table \ref{table:b.outliers} the numbers of upper and lower extreme points and their overall percentage are collected. While deviations from the nominal 95\% coverage are acceptable for transitions with low absolute frequency, such as $(2,3)$, $(3,3)$, $(3,2)$, the remaining transitions require attention. We see that the percentage of outliers higher than the nominal value is mostly due to the upper outliers, which occur as an effect of the difference between the prior opinion on the marginal median of the inter-occurrence times and the median of the observed sequence (compare Table \ref{table:empirical.P}\subref{table:0.median} to Table \ref{table:empirical.P}\subref{table:1.median}). A
few really extreme inter-occurrence times, such as the small values observed at transitions $(1,1)$, $(2,1)$ and the large one at transition $(1,3)$, match unsurprisingly the outlying points in the corresponding qq-plots in Figure~\ref{fig:qqplots}. This fact could be regarded as a lack of fit of the Weibull model, but it could also be due to an imperfect assignment of some events to the macroregion $\text{MR}_3$  or to an insufficient filtering of secondary events (i.e. aftershocks and foreshocks): earthquakes incorrectly assigned to $\text{MR}_3$ and aftershocks or foreshocks can give rise to very short inter-occurrence times; on the other hand, earthquakes which should be in $\text{MR}_3$ but which were attributed to other macroregions can produce very long inter-occurrence times.
\begin{table}[ht]
\begin{center}
\begin{tabular}{llll}
  \hline
 & 1 & 2 & 3 \\ 
  \hline
1 &191 (12) & 172 (18) & 331 (70)\\
2 &214 (22) & 238 (43) & 354 (145)\\
3 & 263 (58) & 203 (55)& 314 (134)\\
 \hline
\end{tabular}
\end{center}
\caption{Predictive means (and standard deviations) of inter-occurrence times for each transition (in days); prior elicited from $\text{MR}_4$.}
\label{table:posterior.time}
\end{table}

\begin{table}[ht]
\begin{center}
\begin{tabular}{cccc|ccc|ccc}
 & \multicolumn{3}{c}{Upper outliers} & \multicolumn{3}{c}{Lower outliers} & \multicolumn{3}{c}{\% of outliers}\\
  \hline
 & 1 & 2 & 3 & 1 & 2 & 3 & 1 & 2 & 3\\ 
  \hline
1 & 8 & 3 & 1 & 2 & 0 & 0 & 15.4\% & 10.0\% & 5.9\%\\
2 & 4& 3& 1& 1& 0 & 0 & 15.6\% &20.0\% & 14.3\%\\
3 & 1 & 0 & 0 & 0 & 1 & 0 & 6.7\% & 11.1\%& 0.0\%\\
 \hline
\end{tabular}
\end{center}
\caption{Number of points having lower or upper posterior $p$-value less than  $2.5$~percent and their percentage; prior elicited from $\text{MR}_4$.}
\label{table:b.outliers}
\end{table}
The shape parameters $\alpha_{ij}$ are particularly important as they reflect an increasing hazard if larger than 1, a decreasing hazard if smaller than 1 and a constant hazard if equal to 1. Table~\ref{table:posterior.alpha-theta}~\subref{table:posterior.alpha} displays the posterior means of these parameters (along with their posterior standard deviations). Finally Table~\ref{table:posterior.P} shows the posterior means of the transition probabilities. Notice that the last row departs from the other two; we will return to this in the following.
\begin{table}[ht]
\centering
    \subtable[Shape parameter $\alpha$   \label{table:posterior.alpha}]
    {\scalebox{1}{
\begin{tabular}{llll}
  \hline
 & 1 & 2 & 3 \\ 
 \hline
	1 & 1.18  (0.06)	& 1.07  (0.08) & 0.94 (0.10) \\ 
  	2 & 1.07 (0.07)	& 0.95 (0.10) & 0.89 (0.14) \\ 
  	3 & 1.04 (0.16)	& 1.03  (0.16)	& 1.11 (0.21)\\ 
   \hline
\end{tabular}
}}\quad
\subtable[Scale parameter $\theta$  \label{table:posterior.theta}]
{\scalebox{1}{
\begin{tabular}{llll}
  \hline
 & 1 & 2 & 3 \\ 
  \hline
1& 201.7 (13.2)& 175.6 (19.1)& 317.8 (67.0)\\
2 & 219.2 (22.5)& 231.0 (40.5)& 327.3 (132.4)\\
3 & 262.7 (57.2) & 201.9 (52.2) & 320.1 (133.5)\\
 \hline
\end{tabular}
}}
\caption{Posterior means (with standard deviations) of the shape parameter $\alpha$ in \subref{table:posterior.alpha} and of the scale parameter $\theta$ in \subref{table:posterior.theta}; prior elicited from $\text{MR}_4$.}
\label{table:posterior.alpha-theta}
\end{table}

\begin{table}[ht]
\begin{center}
\begin{tabular}{rrrr}
  \hline
 & 1 & 2 & 3 \\ 
  \hline
	1 & 0.614 (0.028) & 0.280 (0.026) & 0.106 (0.018)\\ 
  	2 & 0.626 (0.041) & 0.290 (0.038) & 0.085 (0.023)\\ 
  	3 & 0.479 (0.070) & 0.361 (0.067) & 0.160 (0.051)\\ 
   \hline
\end{tabular}
\end{center}
\caption{Summaries of the posterior distributions of the transition matrix $\bp$. Posterior means (with standard deviations); prior elicited from $\text{MR}_4$.}
\label{table:posterior.P}
\end{table}

\begin{figure}[ptbh]
\centering
\includegraphics[scale=0.7]{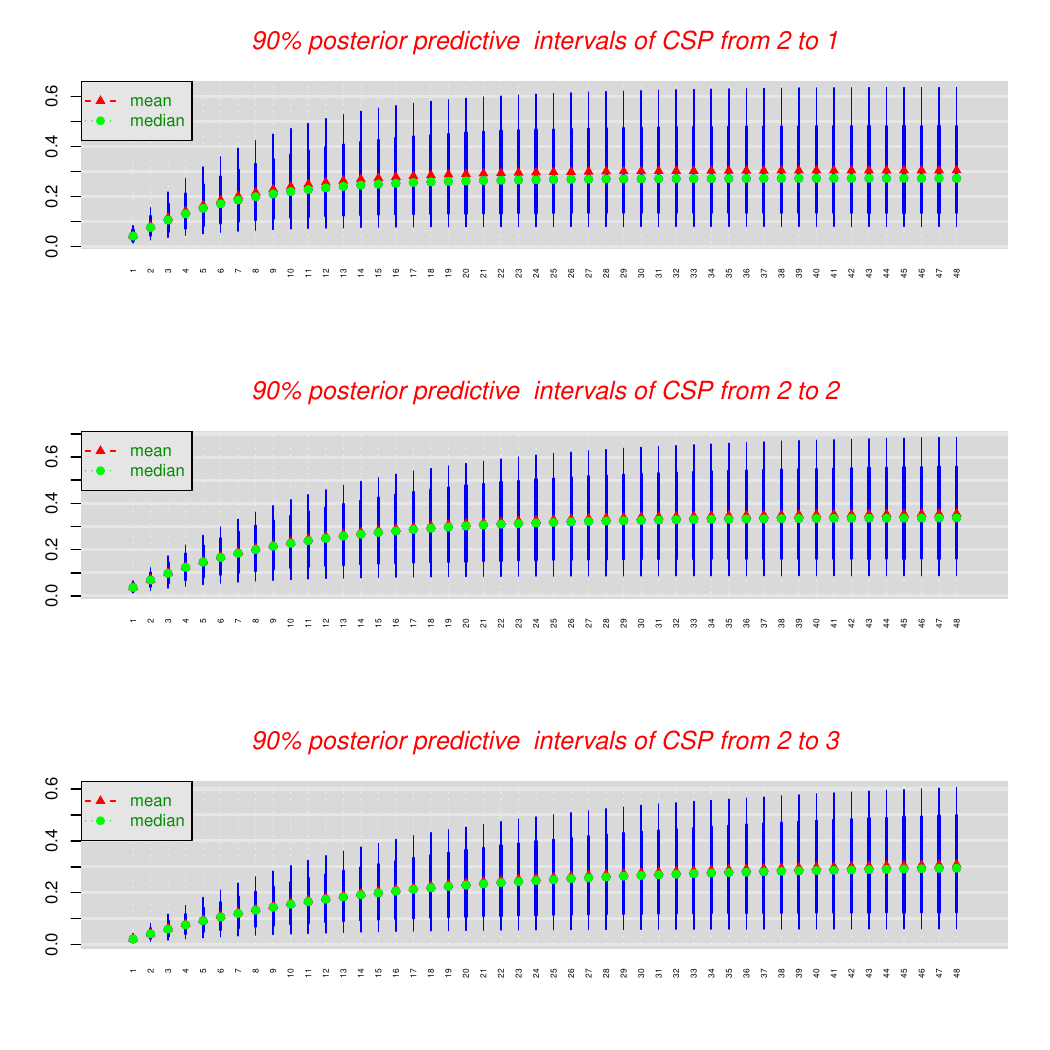}
\caption{Posterior mean and median of CSPs with time origin on 31 December 2002 up to 48 months ahead, along with 90~percent posterior credible intervals. Transitions are from state 2 to state 1, 2 or 3 (first to third panel, respectively).  Months since 31 December 2002 are along the x-axis. The learning set is $MR_4$.}
\label{figure:csp}
\end{figure}

Cross state-probability plots are an attempt at predicting what type of event and when it is most likely to occur. 
A cross state-probability (CSP) $P_{t_0| \Delta x}^{ij}$ represents the probability that the next event will be in state $j$ within a time interval $\Delta x$ under the assumption that the previous event was in state $i$ and $t_0$ time units have passed since its occurrence:
\begin{multline}
	\label{eq:CSP}
 		P_{t_0| \Delta x}^{ij} = P\left(J_{n+1} =j, \ X_{n+1}\le t_0+\Delta x| \ J_n = i, \ X_{n+1} > t_0\right) =  \\
		= \frac{ p_{ij}\left(\bar{F}_{ij}(t_0) - \bar{F}_{ij}(t_0+ \Delta x )\right)}{ \sum_{k \in E} p_{ik} \bar{F}_{ik}(t_0)} \ .
\end{multline}
Figure~\ref{figure:csp} displays the CSPs with time origin on 31 December 2002, the closing date of the \cite{CPTI04} catalogue. At this time, the last recorded event had been in class 2 and had occurred 965 days earlier (so $t_0$ is about 32 months). From these plots we can read out the probability that an event of any given type will occur before a certain number of months. For example, after 24 months, the sum of the mean CSPs in the three graphs indicates that the probability that an event will have occurred is around 88\%, with a larger probability assigned to an event of type 2, followed by type 1 and type 3. The posterior means of the CSPs are also reported in Table~\ref{table:csp}.
\begin{table}[ht]
\begin{center}
	\scalebox{ 0.9}{	\begin{tabular}{rrrrrrrrrrr}
	  \hline
	  &1 Month & 2 Months & 3 Months & 4 Months &5 Months & 6 Months & 1 Year& 2 Years & 3 Years & 4 Years  \\ 
	  \hline
to 1  & 0.045 &   0.080 &   0.113 &   0.140 &   0.164 &   0.184& 0.256 &  0.296 &  0.303&   0.304\\
to 2  & 0.038 &   0.069 &   0.099 &   0.125 &   0.149 &   0.169& 0.257 &  0.327 &  0.348&   0.356\\
to 3  & 0.023 &   0.041 &   0.061 &   0.078 &   0.095 &   0.109& 0.180 &  0.256 &  0.291&   0.309\\
	   \hline
	\end{tabular}}
\end{center}
\caption{CSPs with time origin on 31 December 2002, as represented in Figure~\ref{figure:csp}; prior elicited from $\text{MR}_4$.}
	\label{table:csp}
\end{table}

The predictive capability of our model can be assessed by marking the time of the next event on the relevant CSP plot. In our specific case, the first event in 2003, which can be assigned to the macroregion $\text{MR}_3$ happened in the Forl\`{\i} area on 26 January and was of type 1, with a CSP of 4.5\%. This is a low probability, but a single case is not enough to judge our model, which would be a bad one if repeated comparisons did not reflect the pattern represented by the CSPs. Therefore we repeated the same comparison by re-estimating the model using only the data up to 31 December 2001, 31 December 2000, and so on backwards down to 1992. The results are shown in Table~\ref{table:31.12.92-01}. The boxed numbers correspond to the observed events and it is a good sign that they do not always correspond to very high or very low CSPs, as this would indicate that events occur too late or too early compared to the estimated model. If we were to plot the conditional densities obtained by differentiating the CSPs with respect to $\Delta x$, marking the observed inter-occurrence times on the x-axis, we would observe that very few of them appear in the tails.
%%%%%%
	\begin{table}[h!]
	\centering
	\scalebox{ 0.8}{\begin{tabular}{rrrrrrrrrrrr}
	 & \multicolumn{11}{c}{\textbf{end of catalogue: 31/12/2001;\quad previous event type:2;\quad holding time: 600 days}}\\
	  \hline 
	       & 1 Month & 2 Months & 3 Months & 4 Months & 5 Months & 6 Months & 1 Year & 392 days & 2 Years & 3 Years & 4 Years \\
		%to 1 & 0.053 & 0.095 & 0.135 & 0.169 & 0.199 & 0.224 & 0.318 & \boxed{0.373} & 0.375 & 0.384 & 0.387  \\ 
		to 1 & 0.069 & 0.122 & 0.173 & 0.215 & 0.251 & 0.282 & 0.392 & \boxed{{\color{blue}0.401}} & 0.451 & 0.461 & 0.462 \\
		%to  2 & 0.034 & 0.063 & 0.092 & 0.118 & 0.143 & 0.166 & 0.272 & 0.383 & 0.388 & 0.432 & 0.456 \\ 
		to 2 & 0.038 & 0.068 & 0.097 & 0.123 & 0.146 & 0.166 & 0.248 & 0.256 & 0.310 & 0.328 & 0.333 \\
		%to  3 &  0.017 & 0.031 & 0.043 & 0.054 & 0.063 & 0.070 & 0.098 & 0.115 & 0.115 & 0.119 & 0.121 \\
	   	to 3 & 0.015 & 0.027 & 0.039 & 0.050 & 0.061 & 0.070 & 0.113 & 0.118 & 0.158 & 0.178 & 0.187 \\	
	\hline
	& \multicolumn{11}{c}{\textbf{end of catalogue: 31/12/2000;\quad previous event type: 2;\quad holding time: 235 days}}\\
	  \hline 
	 & 1 Month & 2 Months & 3 Months & 4 Months & 5 Months & 6 Months & 1Year & 2 Years & 757 days & 3 Years & 4 Years \\
	%to 1 &  0.061 & 0.112 & 0.162 & 0.205 & 0.245 & 0.280 & 0.417 & 0.502 & \boxed{0.505} & 0.518 & 0.522 \\ 
	 to 1 &  0.085 & 0.152 & 0.217 & 0.270 & 0.318 & 0.358 & 0.505 & 0.584 & \boxed{{\color{blue}0.586}} & 0.597 & 0.599 	\\
	%to 2 & 0.022 & 0.040 & 0.059 & 0.076 & 0.092 & 0.106 & 0.176 & 0.248 & 0.252 & 0.279 & 0.293 \\ 
	 to 2 & 0.035 & 0.063 & 0.090 & 0.113 & 0.134 & 0.152 & 0.223 & 0.273 & 0.275 & 0.287 & 0.290 \\ 	
	%to 3 & 0.017 & 0.031 & 0.047 & 0.061 & 0.074 & 0.085 & 0.132 & 0.159 & 0.159 & 0.164 & 0.165\\ 
	 to 3 & 0.009 & 0.017 & 0.024 & 0.031 & 0.037 & 0.042 & 0.066 & 0.090 & 0.091 & 0.099 & 0.103\\
	   \hline
	& \multicolumn{11}{c}{\textbf{end of catalogue: 31/12/1999;\quad previous event type: 1;\quad holding time: 177 days}}\\
	  \hline 
	    & 1 Month & 2 Months & 3 Months & 4 Months & 130 days & 5 Months & 6 Months & 1 Year & 2 Years & 3 Years & 4 Years \\
		%to 1 & 0.062 & 0.115 & 0.166 & 0.210 & 0.222 & 0.251 & 0.286 & 0.430 & 0.525 & 0.544 & 0.548 \\ 
		 to 1 & 0.086 & 0.157 & 0.222 & 0.277 & 0.292 & 0.325 & 0.366 & 0.518 & 0.600 & 0.613 & 0.615\\
		%to 2 & 0.038 & 0.074 & 0.110 & 0.142 & \boxed{0.151} & 0.172 & 0.197 & 0.283 & 0.305 & 0.305 & 0.305 \\ 
		 to 2 & 0.035 & 0.063 & 0.090 & 0.113 & \boxed{{\color{blue}0.119}} & 0.133 & 0.151 & 0.220 & 0.269 & 0.281 & 0.284 \\
		%to 3 & 0.013 & 0.023 & 0.033 & 0.042 & 0.045 & 0.050 & 0.058 & 0.091 & 0.122 & 0.135 & 0.140 \\ 
	     to 3 & 0.009 & 0.016 & 0.023 & 0.029 & 0.031 & 0.035 & 0.040 & 0.062 & 0.082 & 0.091 & 0.094 \\  
	\hline	
	& \multicolumn{11}{c}{\textbf{end of catalogue: 31/12/1998;\quad previous event type: 3;\quad holding time: 280 days}}\\
	  \hline 
	   & 1 Month & 2 Months & 3 Months & 4 Months & 5 Months & 6 Months & 188 days & 1 Year & 2 Years & 3 Years & 4 Years \\
	%to 1 &   0.062 & 0.109 & 0.153 & 0.189 & 0.221 & 0.247 & \boxed{0.252} & 0.339 & 0.389 & 0.400 & 0.402  \\ 
	 to 1 &   0.085 & 0.151 & 0.214 & 0.267 & 0.314 & 0.353 & \boxed{{\color{blue}0.361}} & 0.496  & 0.573 & 0.585 &  0.587 \\
	%to 2 &  0.037 & 0.067 & 0.095 & 0.119 & 0.141 & 0.160 & 0.164 & 0.234 & 0.290 & 0.307 & 0.313\\ 
	 to 2 &  0.035 & 0.063 & 0.091 & 0.114 & 0.135 & 0.153 & 0.157 & 0.225 & 0.277 & 0.290 & 0.294 	\\	
	%to 3 & 0.037 & 0.068 & 0.100 & 0.126 & 0.150 & 0.170 & 0.175 & 0.238 & 0.268 & 0.274 & 0.276 \\
     to 3 & 0.010 & 0.018 & 0.025 & 0.032 & 0.039 & 0.045 & 0.046 & 0.071 & 0.096 & 0.106 & 0.111 \\
	   \hline
	& \multicolumn{11}{c}{\textbf{end of catalogue: 31/12/1997;\quad previous event type: 3;\quad holding time: 442 days}}\\
	  \hline 
	   & 1 Month & 2 Months & 85 days & 3 Months & 4 Months & 5 Months & 6 Months & 1 Year & 2 Years & 3 Years & 4 Years \\
	%to 1& 0.074 & 0.131 & 0.176 & 0.184 & 0.227 & 0.264 & 0.294 & 0.402 & 0.463 & 0.475 & 0.479 \\ 
	 to 1& 0.078 & 0.138 & 0.187 & 0.196 & 0.244 & 0.286 & 0.320 & 0.448 & 0.516 & 0.526 & 0.528 \\
	%to 2 & 0.054 & 0.096 & 0.131 & 0.138 & 0.173 & 0.205 & 0.232 & 0.343 & 0.428 & 0.455 & 0.466 \\ 
	 to 2 & 0.036 & 0.065 & 0.090 & 0.094 & 0.118 & 0.140 & 0.159 & 0.237 & 0.294 & 0.310 & 0.315\\
	%to 3 & 0.010 & 0.016 & \boxed{0.021} & 0.021 & 0.025 & 0.028 & 0.030 & 0.037 & 0.041 & 0.042 & 0.043 \\
	 to 3 & 0.012 & 0.022 & \boxed{{\color{blue}0.030}} & 0.031 & 0.040 & 0.048 & 0.056 & 0.090 & 0.123 & 0.138 & 0.145\\
	   \hline 
& \multicolumn{11}{c}{\textbf{end of catalogue: 31/12/1996;\quad previous event type: 3;\quad holding time: 77 days}}\\
	  \hline
	 & 1 Month & 2 Months & 3 Months & 4 Months & 5 Months & 6 Months & 1 Year & 450 days & 2 Years & 3 Years & 4 Years \\ 
	  %to 1 & 0.073 & 0.131 & 0.186 & 0.231 & 0.271 & 0.304 & 0.422 & 0.447 & 0.483 & 0.493 & 0.496 \\ 
	   to 1 & 0.084 & 0.152 & 0.218 & 0.273 & 0.323 & 0.366 & 0.525 & 0.562 & 0.614 & 0.628 & 0.630 \\  
	  %to 2 & 0.043 & 0.075 & 0.106 & 0.132 & 0.155 & 0.174 & 0.248 & 0.268 & 0.300 & 0.314 & 0.319  \\ 
	   to 2 & 0.036 & 0.063 & 0.090 & 0.113 & 0.133 & 0.150 & 0.219 & 0.236 & 0.266 & 0.277 & 0.281 \\
	  %to 3 & 0.012 & 0.027 & 0.049 & 0.076 & 0.105 & 0.129 & 0.172 & \boxed{0.175} & 0.178 & 0.179 & 0.179\\
	   to 3 & 0.008 & 0.015 & 0.022 & 0.027 & 0.032 & 0.037 & 0.056 &\boxed{{\color{blue}0.062}} & 0.074 & 0.081 & 0.084 \\
	   \hline
	& \multicolumn{11}{c}{\textbf{end of catalogue: 31/12/1995;\quad previous event type: 1;\quad holding time: 3100 days}}\\
	  \hline
	  & 1 Month & 2 Months & 3 Months & 4 Months & 5 Months & 6 Months & 288 days & 1 Year & 2 Years & 3 Years & 4 Years \\ 
	 %to 1 &0.173 & 0.304 & 0.420 & 0.511 & 0.589 & 0.651 & 0.798 & 0.861 & 0.964 & 0.982 & 0.985 \\ 
	  to 1 &0.004 & 0.008 & 0.011 & 0.014 & 0.017 & 0.019 & 0.024 & 0.027 & 0.033 & 0.035 & 0.035 \\ 	
	%to 2 &  0.001 & 0.002 & 0.002 & 0.003 & 0.003 & 0.004 & 0.004 & 0.004 & 0.005 & 0.005 & 0.005 \\ 
	  to 2 & 0.022 & 0.041 & 0.059 & 0.075 & 0.091 & 0.104 & 0.143 & 0.165 & 0.222 & 0.244 & 0.253\\	
	 %to 3 & 0.001 & 0.003 & 0.004 & 0.004 & 0.005 & 0.005 & \boxed{0.007} & 0.007 & 0.008 & 0.008 & 0.008 \\
	  to 3 & 0.042 & 0.079 & 0.115 & 0.148 & 0.180 & 0.208 & \boxed{{\color{blue}0.296}} & 0.348 & 0.509 & 0.590 & 0.633 \\
	   \hline
	& \multicolumn{11}{c}{\textbf{end of catalogue: 31/12/1994;\quad previous event type: 1;\quad holding time: 2735 days}}\\
	  \hline
	  & 1 Month & 2 Months & 3 Months & 4 Months & 5 Months & 6 Months & 1 Year & 653 days & 2 Years & 3 Years & 4 Years \\
	%to 1& 0.169 & 0.295 & 0.410 & 0.502 & 0.580 & 0.643 & 0.857 & 0.954 & 0.964 & 0.982 & 0.986 \\ 
	 to 1& 0.005 & 0.010 & 0.014 & 0.018 & 0.021 & 0.023 & 0.034 & 0.041 & 0.041 & 0.043 & 0.043\\
	%to  2 & 0.001 & 0.002 & 0.002 & 0.003 & 0.003 & 0.004 & 0.005 & 0.005 & 0.005 & 0.005 & 0.005 \\ 
	  to 2 & 0.024 & 0.043 & 0.062 & 0.080 & 0.096 & 0.110 & 0.174 & 0.225 & 0.233 & 0.256 & 0.265\\
	%to  3 & 0.001 & 0.002 & 0.003 & 0.004 & 0.004 & 0.005 & 0.007 & \boxed{0.007} & 0.008 & 0.008 & 0.008 \\
	 to  3 & 0.041 & 0.076 & 0.112 & 0.144 & 0.175 & 0.204 & 0.341 & \boxed{{\color{blue}0.473}} & 0.497 & 0.575 & 0.617\\		  
	 \hline
& \multicolumn{11}{c}{\textbf{end of catalogue: 31/12/1993;\quad previous event type: 1;\quad holding time: 2370 days}}\\
	  \hline
	  & 1 Month & 2 Months & 3 Months & 4 Months & 5 Months & 6 Months & 1 Year & 2 Years & 1018 days & 3 Years & 4 Years \\
	%to 1 & 0.166 & 0.289 & 0.403 & 0.495 & 0.573 & 0.636 & 0.854 & 0.964 & 0.981 & 0.982 & 0.986  \\ 
	 to 1 & 0.007 & 0.013 & 0.019 & 0.024 & 0.028 & 0.031 & 0.045 & 0.055 & 0.056 & 0.057 & 0.057 \\	
	%to 2 & 0.001 & 0.002 & 0.003 & 0.003 & 0.004 & 0.004 & 0.005 & 0.005 & 0.005 & 0.005 & 0.005 \\ 
     to 2 & 0.025 & 0.046 & 0.067 & 0.086 & 0.103 & 0.118 & 0.186 & 0.248 & 0.268 & 0.271 & 0.280\\
	%to 3 & 0.001 & 0.002 & 0.003 & 0.003 & 0.004 & 0.004 & 0.006 & 0.007 & \boxed{0.007} & 0.007 & 0.007  \\
	 to 3 & 0.040 & 0.074 & 0.108 & 0.140 & 0.170 & 0.197 & 0.329 & 0.479 & \boxed{{\color{blue}0.542}} & 0.554 & 0.593 \\
	   \hline
	& \multicolumn{11}{c}{\textbf{end of catalogue: 31/12/1992;\quad previous event type: 1;\quad holding time: 2005 days}}\\
	  \hline
	  & 1 Month & 2 Months & 3 Months & 4 Months & 5 Months & 6 Months & 1 Year & 2 Years & 3 Years & 1383 days & 4 Years \\ 
	 %to 1  & 0.161 & 0.283 & 0.395 & 0.486 & 0.564 & 0.628 & 0.849 & 0.963 & 0.983 & 0.986 & 0.987 \\ 
	  to 1  & 0.011 & 0.019 & 0.027 & 0.034 & 0.040 & 0.045 & 0.065 & 0.078 & 0.080 & 0.081 & 0.081\\ 
	 %to 2 & 0.001 & 0.002 & 0.003 & 0.003 & 0.004 & 0.004 & 0.005 & 0.006 & 0.006 & 0.006 & 0.006 \\
	  to 2 & 0.028 & 0.051 & 0.074 & 0.094 & 0.113 & 0.130 & 0.202 & 0.267 & 0.290 & 0.298 & 0.299\\
	 %to 3 & 0.001 & 0.002 & 0.002 & 0.003 & 0.003 & 0.004 & 0.005 & 0.006 & 0.006 & \boxed{0.006} & 0.006 \\
	  to 3 & 0.038 & 0.070 & 0.103 & 0.132 & 0.161 & 0.186 & 0.311 & 0.451 & 0.520 & \boxed{{\color{blue}0.551}} &  0.557 \\
	   \hline
	\end{tabular}}
	\caption{CSPs as the end of the catalogue shifts back by one-year steps. The numbers in boxes are the probability that the next observed event has occurred at or before the time when it occurred and is of the type that has been observed. The prior was elicited from $\text{MR}_4$.}
	\label{table:31.12.92-01}
	\end{table}

The examination of the posterior distributions of transition probabilities and of the predictive distributions of the inter-occurrence times can give some insight into the type of energy release and accumulation mechanism. We consider two mechanisms, the time predictable model (TPM) and the slip predictable model (SPM).

In the TPM, it is assumed that when a maximal energy threshold is reached, some fraction of it (not always the same) is released and an earthquake occurs. The consequence is that the time until the next earthquake increases with the amplitude of the last earthquake. So, the holding time distribution depends on the current event type, but not on the next event type, that is, we expect $F_{ij}(t)=F_{i\cdot}(t)$, $j=1,2,3$. The strength of an event does not depend on the strength of the previous one, because every time the same energy level has to be reached for the event to occur. So we expect $p_{ij}=p_{\cdot j}$, $j=1,2,3$, that is, a transition matrix with equal rows. If this is the case, the CSPs~\eqref{eq:CSP} would simplify as follows,
\begin{equation}\label{eq:CSPTPM}
P_{t_0|\Delta x}^{ij} = \frac{ p_{ij}\left(\bar{F}_{ij}(t_0) - \bar{F}_{ij}(t_0+ \Delta x )\right)}{ \sum_{k \in E} p_{ik} \bar{F}_{ik}(t_0)}
= \frac{p_{\cdot j} \left(\bar{F}_{i\cdot}(t_0) - \bar{F}_{i\cdot}(t_0+ \Delta x )\right)}{\bar{F}_{i\cdot}(t_0)} \ ,
\end{equation}
so that, under the TPM assumption,  given $i$, they are proportional to each other as $j=1,2,3$ for any $\Delta x$, and the ratio
$P_{t_0|\Delta x}^{ij}/P_{t_0|\Delta x}^{ik}$ equals $p_{\cdot j}/p_{\cdot k}$ for any pair $(j,k)$.

In the SPM, after an event, energy falls to a minimal threshold and increases until the next event, where it starts to increase again from the same threshold. The consequence is that the energy of the next earthquake increases with time since the last earthquake. So, the magnitude of an event depends on the length of the holding time, but not on the magnitude of the previous one, because energy always accumulates from the same threshold. In this case again $p_{ij} = p_{\cdot j}$, but $F_{ij}(t) = F_{\cdot j}(t)$, so 
\begin{equation}
	\label{eq:CSPSPM}
P_{t_0|\Delta x}^{ij} = 
\frac{ p_{\cdot j}\left(\bar{F}_{\cdot j}(t_0) - \bar{F}_{\cdot j}(t_0+ \Delta x )\right)}{ \sum_{k \in E} p_{\cdot k} \bar{F}_{\cdot k}(t_0)} \ .
\end{equation}
Then,  under the SPM assumption,  CSPs are equal to each other as $i=1,2,3$ for any $\Delta x$, given $j$.

An additional feature that can help discriminate between the TPM and the SPM is the tail of the holding time distribution: for a TPM, the tail of the holding time distribution is thinner \textit{after} a weak earthquake than \textit{after} a strong one; for an SPM, the tail of the holding time is thinner \textit{before} a  weak earthquake than \textit{before} a strong one. 

In the present case the posterior mean of the third row of $\textbf{p}$, see Table \ref{table:posterior.P}, is clearly different from the other two rows, unlike the empirical transition matrix derived from Table \ref{table:empirical.P}\subref{table:1.empirical.P}, because of the prior information from $\text{MR}_4$. So, with this prior, both the TPM and the SPM are excluded.

On the other hand, things change with the noninformative prior elicited without a learning set.
In this case, we let all the Dirichlet hyperparameters $\g_{ij} $'s  be equal to 2. Following Section~ \ref{sec:scarce}, the missing learning  set for each string $(i,j)$ is substituted by a unique fictitious observation $\tilde{t}^{ij}_{0.5}$ uniformly distributed over $(1, 5000)$ days, and $m_{ij}$ is set to one; this establishes the prior for $\theta_{ij}$. 
The prior of $\alpha_{ij}$ derived from Equation \eqref{pi-3} with $m_{ij}=1$ and $c_{ij}=2$ (taken from Table~\ref{table:hyperparameters}) is
$$
\pi_3(\alpha_{ij}) \propto \left(1-\frac{\alpha_{ij}}{\alpha_{0,ij}}\right) \indic(\alpha_{0,ij}\le\alpha_{ij}\le\alpha_{1,ij})
$$
with $\alpha_{0,ij}=2/(2+N_{ij})$ (see Table \ref{table:empirical.P}\subref{table:1.empirical.P} for the $N_{ij}$'s) and $\alpha_{1,ij}=10$.  Note that on our current sample,  the lower limit $\alpha_{0,ij}$ is always smaller than $2/3$. %  suggested by Table~\ref{table:hyperparameters}, 
%Besides, experiments showed that $2/3$ deteriorates the predictive performance of the model, by introducing prior information inconsistent with the present dataset.

With this prior specification, the posterior distributions of the rows of the transition matrix do not differ significantly, as seen from Table~\ref{table:lessinformative.posterior.P}, so we can assume $p_{ij}=p_{\cdot j}$ for all indexes $i$ and examine the ratios of CSPs to verify the TPM and the SPM hypotheses.
\begin{table}[ht]
\begin{center}
\begin{tabular}{rrrr}
  \hline
 & 1 & 2 & 3 \\ 
  \hline
1 & 0.569 (0.046) & 0.271 (0.041)& 0.160 (0.033)\\
2 & 0.568 (0.064)& 0.283 (0.059)& 0.149 (0.046)\\
3 & 0.500 (0.085)& 0.323 (0.078)& 0.177 (0.065)\\
   \hline
\end{tabular}
\end{center}
\caption{Posterior means (with standard deviations) of the transition matrix $\bp$ with the noninformative prior.}
\label{table:lessinformative.posterior.P}
\end{table}

\begin{figure}[htbp]
	\includegraphics[scale=0.5]{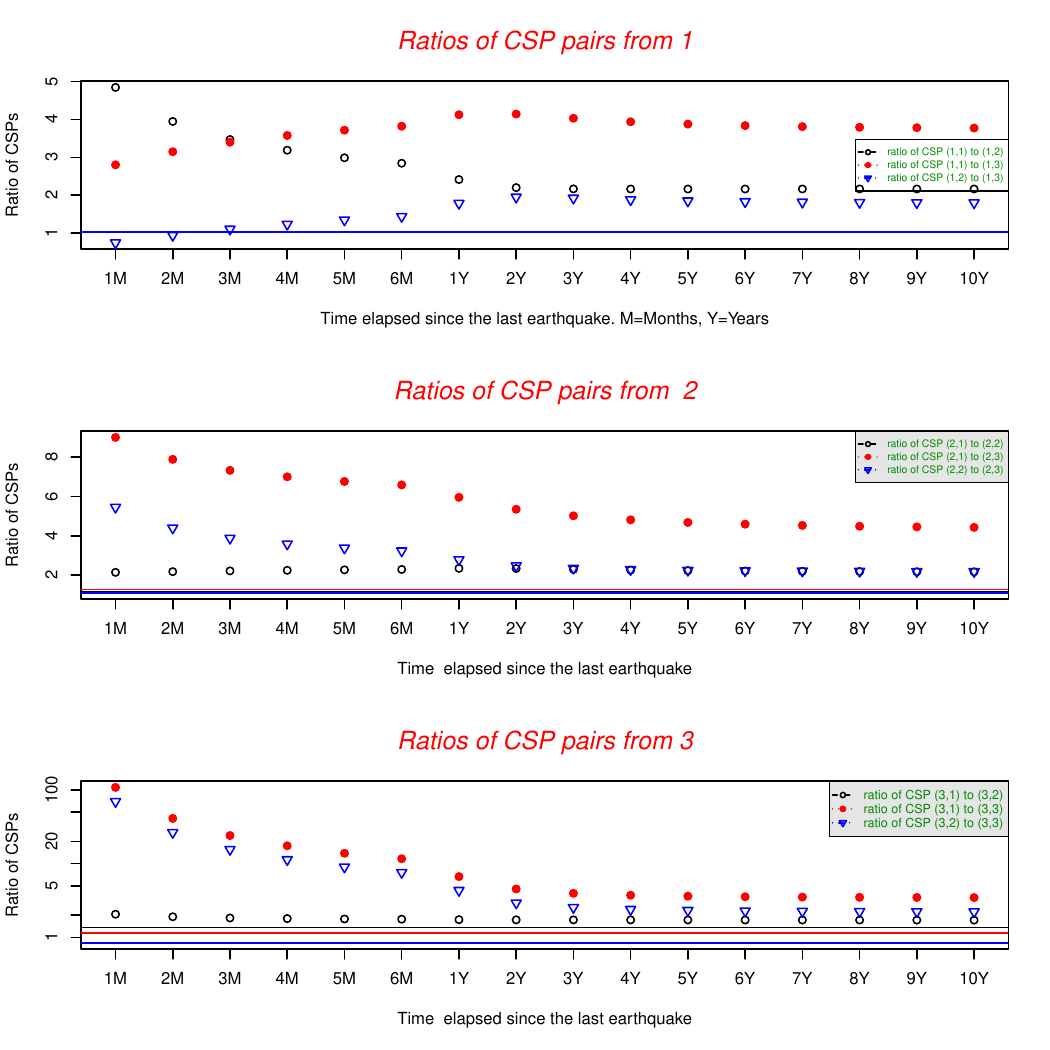}
	\caption{Checking the TPM: posterior means of ratios of CSPs $P_{0|\Delta x}^{ij}/P_{0|\Delta x}^{ik}$, with time origin at 0, up to 10 years ahead. Transitions are from state 1, 2 and 3 to state 1, 2 or 3. Horizontal lines indicate the theoretical values of the ratios for the TPM. The prior distribution is noninformative. }
	\label{figure:improper_ratio_TPM}
\end{figure}

\begin{figure}[htbp]
	\includegraphics[scale=0.5]{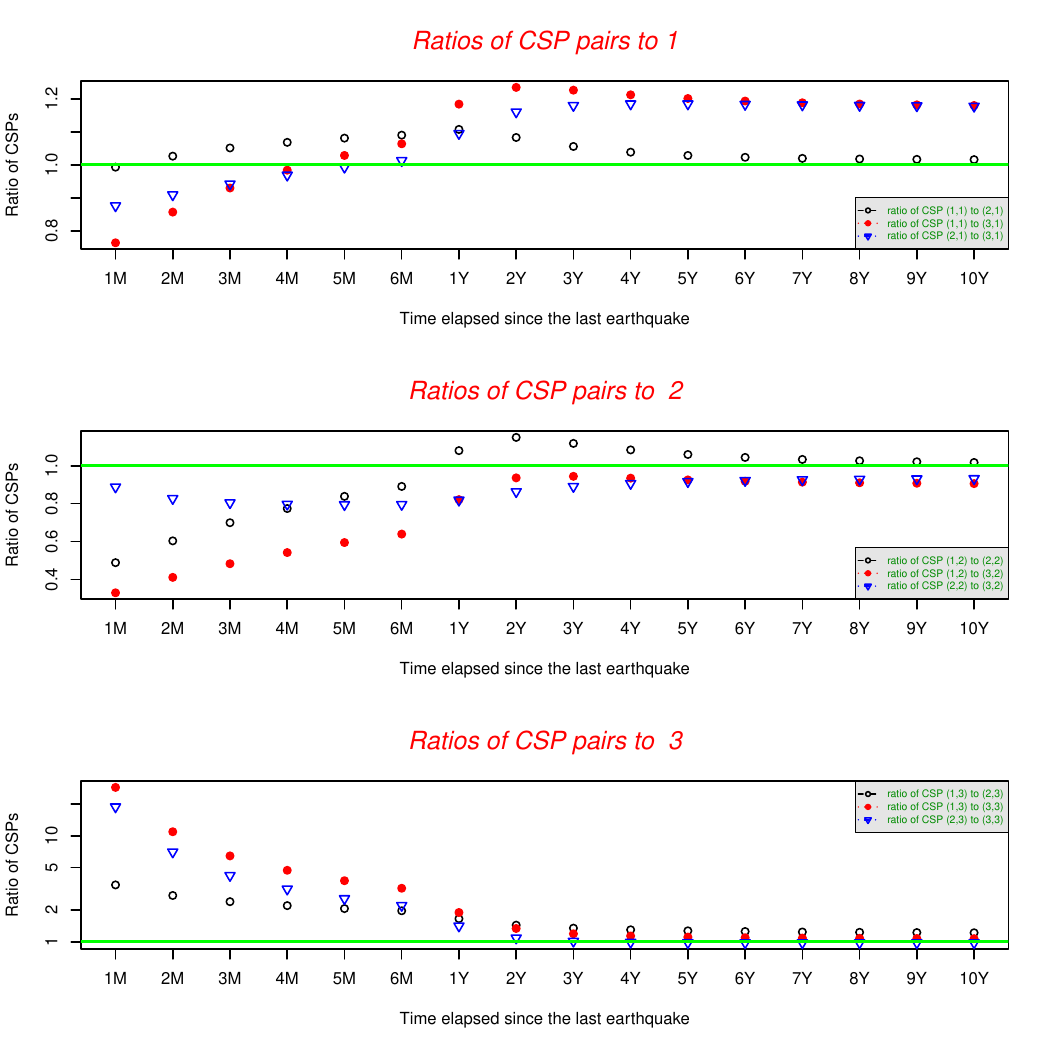}
	\caption{Checking the SPM: posterior means of ratios of CSPs $P_{0|\Delta x}^{ij}/P_{0|\Delta x}^{kj}$, with time origin at 0, up to 10 years ahead. Transitions are from state 1, 2 and 3 to state 1, 2 or 3. Green horizontal lines at one indicate the theoretical value of the ratios for the SPM. The prior distribution is noninformative. }
	\label{figure:improper_ratio_SPM}
\end{figure}

Figures~\ref{figure:improper_ratio_TPM} and \ref{figure:improper_ratio_SPM} display the posterior means of the ratios of the CSPs as a function of $\Delta x$ for $t_0=0$, with the noninformative prior. For the TPM the generic ratio of two CSPs indexed by $(i,j)$ and $(i,k)$ should be approximately constant and close to $p_{\cdot j}/p_{\cdot k}$, where the $p_{\cdot j}$ represents the common values of the entries in the $j$-th column of $\bp$, under the TPM. The horizontal lines in Figure~\ref{figure:improper_ratio_TPM} are the posterior expectations of $p_{ij}/p_{ik}$, which would estimate $p_{\cdot j}/p_{\cdot k}$ if the TPM assumption were true. For the SPM, the ratio of CPSs, now indexed by $(i,j)$ and $(k,j)$, should be close to one. The plots indicate that it is not so, therefore neither the TPM nor the SPM are supported by the data.

As for the TPM, this finding is confirmed by the examination of the posterior probabilities that $\a_{ij}<\a_{ik}$ and $\theta_{ij}<\theta_{ik}$, for any given $i$ and $j\ne k$: $\Pr(\a_{ij}<\a_{ik}|\bj, \bx, u_T)$ is 0.55 for string $(2,1)$ versus string $(2,3)$ and 0.51 for $(3,1)$ versus $(3,2)$, but it is either larger than 0.75 or smaller than 0.35 for all the other strings; $\Pr(\theta_{ij}<\theta_{ik}|\bj, \bx, u_T)$ is 0.61 for $(1,2)$ versus $(1,3)$ and is 0.53 for $(3,1)$ versus $(3,2)$, but it is lower than 0.39 for all the other strings. As for the SPM, we have examined $\Pr(\a_{ij}<\a_{kj}|\bj, \bx, u_T)$ and  $\Pr(\theta_{ij}<\theta_{kj}|\bj, \bx, u_T)$ for any $j$ and $i\ne k$: $\Pr(\a_{ij}<\a_{kj}|\bj, \bx, u_T)$ is 0.51 for $(2,2)$ versus $(3,2)$, but is either larger than 0.75 or smaller than 0.35 for all the other strings; $\Pr(\theta_{ij}<\theta_{kj}|\bj, \bx, u_T)$ is between 0.44 and 0.63 for three comparisons but is either larger than 0.73 or smaller than 0.30 for the remaining ones.

\section{Concluding remarks}
\label{sec:6}
We have presented a complete Bayesian methodology for the inference on semi-Markov processes, from the elicitation of the prior distribution, to the computation of posterior summaries, including a guidance for its JAGS implementation.
In particular, we have examined in detail the elicitation of the joint prior density of the shape and scale parameters of the Weibull-distributed holding times (conditional on the transition between two given states), deriving a specific class of priors in a natural way, along with a method for the determination of hyperparameters based on ``learning data'' and moment existence conditions. This framework has been applied to the analysis of seismic data, but it can be adopted for inference on any system for which a Markov Renewal process is plausible. A possible and not-yet explored application is the modelling of voltage sags (or voltage dips) in power engineering: the state space would be formed by different classes of voltage, starting from voltage around its nominal value, down to progressively deeper sags. In the engineering literature, the dynamic aspect of this problem is in fact disregarded, while it could help bring additional insight into this phenomenon.

With regard to the seismic data analysis, other uses of our model can be envisaged. The model can be applied to areas with a less complex tectonics, such as Turkey, by replicating for example Alvarez's analysis. Outliers, such as those appearing in Figure \ref{figure:caterpillar_times}, could point at events whose assignment to a specific seismogenic source should be re-discussed. The analysis of earthquake occurrence can support decision making related to the risk of future events. We have not examined this issue here, but a methodology is outlined by \cite{Cano}.

A final note concerns the more recent Italian seismic catalogue \cite{CPTI11}, including events up to the end of 2006. Every new release of the catalogue involves numerous changes in the parameterization of earthquakes; as the DISS event classification by macroregion is not yet available for events in this catalogue we cannot use this more recent source of data.

\appendix
\section{Gibbs sampling}
\label{sec:gibbsampling}
Here we derive the full conditional distributions involved in the Gibbs sampling and give indications on its JAGS implementation.
\subsection{Full conditional distributions}
Let the last holding time be censored i.e. $u_T>0$.  Hence,  
in  order to obtain some simple  full conditional distributions and then an efficient Gibbs sampling, we introduce the auxiliary variable $j_{\t +1}$
which represents the unobserved state following the last visited state $j_\t$. Moreover, let 
$\bt_{\bq}  = (t^{ij}_{q_{ij}}, \;\; i,j\in E)$.  Each hyperparameter $t^{ij}_{q_{ij}}$ may be either a known constant or $t^{ij}_{q_{ij}}$ is  uniformly distributed  over an interval $(t_1, t_2)$. Moreover, all of them are independent of each other.
Thus the state space of the Gibbs sampler is $(\bp, \ba, \bth,j_{\t+1}, \bt_{\bq})$ and the following full likelihood  derived from \eqref{eq:likelihood-param}:
\begin{multline}
	\label{eq:full-likelihood-param}
 L(\bj, \bx, u_T , j_{\t+1}|\bp, \ba, \bth) =\prod_{i,k\in E}p_{ik}^{N_{ik}}\times \\
\times  \prod_{i,k\in E}\left[\a_{ik}^{N_{ik}}\frac{1}{\theta_{ik}^{\a_{ik}N_{ik}}}\left(\prod_{\r =1}^{N_{ik}}x^{\r}_{ik}\right)^{\a_{ik}-1} \times 
\exp\left\{-\frac{1}{\theta_{ik}^{\a_{ik}}}\sum_{\r  =1}^{N_{ik}}(x^{\r}_{ik})^{\a_{ik}}\right\}\right]\times \\
 \times \left(p_{j_{\t} j_{\t+1}}
\exp\left\{-\left(\frac{u_T}{\theta_{j_{\t} j_{\t+1}}}\right)^{\a_{j_{\t}  j_{\t+1}}}\right\}\right)\ 
\end{multline}
is multiplied by the prior and used to determine the full conditionals.
For every $i$ and $j$ let
\begin{align*}
\bp_{(-i)}   & = \mbox{the transition matrix  $\bp$ without the $i$-th row} ,  \\
\ba_{(-ij)}   & = (\a_{hk}, \;\; h,k\in E,\quad (h,k)  \neq (i,j) ) \ , \\
\bth_{(-ij)}  & = (\th_{hk}, \;\; h,k\in E,\quad (h,k) \neq (i,j) ) \ , \\
\bt_{{\bq}(-ij)}  & =   (t^{hk}_{q_{hk}}, \;\; h,k\in E,\quad (h,k)  \neq (i,j) ) , \\
\tilde{N}_{ij} & = {N}_{ij} + \indic{\left((j_{\t},j_{\t+1})=(i,j)\right)},  \quad
\tilde{\mathbf N}_{i}  =\left(\tilde{N}_{ij}, \;\; j=1,\ldots, s \right),\\
\tilde{M}_{ij}(\a_{ij}) & = \sum_{\rho=1}^{{N}_{ij}}(x_{ij}^{\rho})^{\a_{ij}} + \ u_{T}^{\a_{ij}}\indic( (j_{\t},j_{\t+1})=(i,j)), \\
{\C}_{ij}  & = \prod_{\rho=1}^{{N}_{ij}}x_{ij}^{\rho} \ . 
\end{align*}
%Furthermore,  %denote by ${{\bf C}}$ and ${ \tilde{\bf M}}$ the matrices ${ {\bf C}}=(\C_{ij})_{i,j\in E}$ and ${\tilde{\bf M}}=(M_{ij}(\a_{ij}))_{i,j\in E}$. 
%let   $m_{ij}$ be the number of visits to the string $(i,j)$ in the learning  dataset, $y_{ij}^{\r}$  the inter-occurrence time at $\r$th visit to $(i,j)$, 
%for $\r=1,\dots m_{ij}$,  and  ${\bf y}_{m_{ij}}=(y_{ij}^{1},\dots ,y^{m_{ij}}_{ij})$.  \newline \indent 
The following result on the  full conditional distributions of the Gibbs sampling holds.
\begin{proposition}
	\label{prop:A.1}
	Let the prior on $(\bp,\ \ba, \ \bth, \  \bt_{\bq})$ be the following %given as in Section~\ref{sec:3}, i.e. 
\begin{enumerate}[i)] 
	\item $\bp$ is independent of $\ba$ and $\bth$ and the rows of $\bp$ are $s$ independent vectors with Dirichlet distribution with parameters $\bg_1, \cdots ,\bg_s$ and total mass $c_1, \cdots c_s$, respectively,
	\item the $\th_{ij}$'s, given the $\a_{ij}$'s and the $t^{ij}_{q_{ij}}$'s, are independent with $\th_{ij}|\a_{ij} \sim {\mathcal GIG}(m_{ij}, b_{ij}(t^{ij}_{q_{ij}},\a_{ij}), \a_{ij})$, where 
	\begin{equation*}
		%	\label{eq:b_q^*}
		b_{ij}(t^{ij}_{q_{ij}},\a_{ij}) = \left(t^{ij}_{q_{ij}}\right)^{\a_{ij}}[(1-q_{ij})^{-1/m_{ij}}-1]^{-1} \ , 
	\end{equation*}
and  $t^{ij}_{q_{ij}}$ is either a known constant or $t^{ij}_{q_{ij}}$ is  uniformly distributed  over $(t_1, t_2)$,
	\item $\pi_{3,ij}(\a_{ij})\propto \a_{ij}^{m_{ij}-c_{ij}}\left(\a_{ij}-\a_{0,ij}\right)^{c_{ij}-1}\exp\{-m_{ij}d_{ij}\a_{ij}\}\indic(\a_{ij}\in {I_{ij}})$,  
	$m_{ij}>0,  c_{ij}>0$,  $\a_{0,ij}>0 \mbox{ and }   d_{ij}\ge0$ where
	$$
	I_{ij} = 
	\begin{cases} 
		(\a_{0,ij}, \a_{1,ij}) & \mbox{ if } d_{ij} = 0\\ 
		(\a_{0,ij}, \infty) & \mbox{ if } d_{ij} >  0 \ .
		\end{cases}
	$$ 
\end{enumerate}
Then
\begin{enumerate}[$a)$]
	\item the conditional distribution of $\bp_i$, given $\bj, \ \bx, \ u_T, \ j_{\t+1},\ \bp_{(-i)}, \ \ba, \ \bth$ and $\bt_{\bq}$ is a Dirichlet distribution with parameter $\tilde{\mathbf N}_{i}+\bg_i$;
	\item the conditional distribution of $\th_{ij}^{\a_{ij}}$, given $\bj, \ \bx , \ u_T, \ j_{\t+1},  \ \bp ,  \ \ba, \ \bth_{(-ij)}$ and $\bt_{\bq}$  is an inverse Gamma distribution with shape $m_{ij}+{N}_{ij}$ and  rate $b_{ij}(t^{ij}_{q_{ij}},\a_{ij})+\tilde{M}_{ij}(\a_{ij})$;
	\item the conditional density of $\a_{ij}$, given $\bj ,   \ \bx, \  u_T, \ j_{\t+1},  \ \bp,  \ \ba_{(-ij)}, \ \bth$ and  $\bt_\bq$ is proportional to 
\begin{multline}
\label{eq:pi3_posterior}
\a_{ij}^{N_{ij}+1+m_{ij}-c_{ij}}\left(\a_{ij}-\a_{0,ij}\right)^{c_{ij}-1} 
	\times 
		\exp\left\{ - \left(m_{ij}d_{ij} -\log \frac{\C_{ij} (t^{ij}_{q_{ij}})^{m_{ij}}}{\th_{ij}^{N_{ij} + m_{ij}}}\right)\a_{ij}\right\} 
	\times \\ 
	\times \exp\left\{- \frac{b_{ij}(t^{ij}_{q_{ij}},\a_{ij})+\tilde{M}_{ij}(\a_{ij})  }{\th_{ij}^{\a_{ij}}}\right\} \indic(\a_{ij}\in {I_{ij}})\ ,
\end{multline}
and it is log-concave if $c_{ij}\ge 1$; %$m_{ij}>0,  c_{ij}\ge 1,  \a_{0,ij} > 0$ and $d_{ij}>0$. 
%where $t^{ij}_{q_{ij}}$ is determined from Table~\ref{table:hyperparameters};
\item  the conditional density of the unseen state $J_{\t+1}$, given $\bj, \ \bx, \   u_T, \  \bp, \ \ba,\bth$ and $\bt_{\bq}$,  is 
\begin{equation*}
%P(J_{\t+1}=j \ |\ \bj, \bx, u_T , \bp, \ba, \bth ) = 
 \frac{p_{j_{\t}j}\exp\left\{-\left(\frac{u_T}{\th_{j_{\t}j}}\right)^{\a_{j_{\t}j}}\right\}}{\sum_{k\in E}
p_{j_{\t}k}\exp\left\{-\left(\frac{u_T}{\th_{j_{\t}k}}\right)^{\a_{j_{\t}k}}\right\}} \, ; 
\end{equation*}
\item if $t^{ij}_{q_{ij}}$ is  uniformly distributed  over  $(t_1, t_2)$, then  its conditional distribution  given $\bj, \ \bx, \   u_T, \  \bp, \ \ba, \ \bth$ and $\bt_{{\bq}(-ij)}$ is a doubly-truncated at $(t_1, t_2)$ generalized Gamma with parameters 
$a= \th_{ij}[(1-q_{ij})^{-1/m_{ij}}-1]^{1 /\a_{ij}}$, $d=\a_{ij}m_{ij}+1$  and $p=\alpha_{ij}$, i.e. 
$$
\pi(t_{q_{ij}}^{ij} | \bj, \bx,    u_T,   \bp,  \ba,  \bth, \bt_{{\bq}(-ij)}) =
\frac{p/a^d}{\Gamma(d/p)} \left(t_{q_{ij}}^{ij}\right)^{d-1} \exp\left\{-\left(t_{q_{ij}}^{ij} / a \right)^p \right\} \indic(t_1 <t^{ij}_{q_{ij}}< t_2) . 
$$
 %The truncation points are given by the smallest and the largest inter-occurrence times included in all the learning dataset.
\end{enumerate}
\end{proposition}
\begin{proof}
As the row $\bp_{i}$ is  independent of  $(\bp_{(-i)},\ba,\bth, \bt_{\bq})$,  conditionally on the data and $j_{\t+1}$, then 
\begin{equation*}
	\pi(\bp_i|\bj, \bx , u_T, \ j_{\t+1} ,\bp_{(-i)},  \ba,\bth,\bt_{\bq})	\propto  L(\bj, \bx, u_T, \ j_{\t+1}| \bp, \ba,\bth) 
	%\times L(\bj^*, \bx^*|j^h_{\t(h)}: \; h=1,\ldots, d, \bth,\ba,\bp) 
	\times \pi_{1,i}(\bp_i) \propto  \prod_{j\in E} p_{ij}^{\tilde{N}_{i,j}} \times \prod_{j\in E} p_{ij}^{\gamma_{ij}-1} ,
\end{equation*}
where $\pi_{1,i}$ denotes the Dirichlet prior of $\bp_i$. Hence point $a)$ of Proposition \ref{prop:A.1} follows. 

As regards the full conditional distribution of $\theta_{ij}$, % given $(\bj, \bx , u_T, \ j_{\t+1},\bp,  \ba,\bth_{(-ij)},\bt_{\bq})$, 
we have 
\begin{align*}
	  %\pi_{2,ij}
		\pi(\th_{ij}  & \ |\  \bj, \bx , u_T, \ j_{\t+1}, \bp,  \ba,\bth_{(-ij)},\bt_{\bq})  
	 \propto  L(\bj, \bx, u_T, \ j_{\t+1} \ |\bp,  \ba,\bth) \times \pi_{2,ij}(\th_{ij}|\alpha_{ij}, t^{ij}_{q_{ij}}) \\
	& \propto     
		\prod_{i,k\in E} \left[\a_{ik}^{N_{ik}}\frac{\C_{ik}^{\a_{ik}-1}}{\theta_{ik}^{\a_{ik}N_{ik}}} 	
		\times\exp\left\{-\frac{\tilde{M}_{ik}(\a_{ik})}{\theta_{ik}^{\a_{ik}}}\right\} 	\right]
		%\times 	\frac{\a_{ij}\left(t^{ij}_{q_{ij}}\right)^{m_{ij}\a_{ij}}}{\Gamma(m_{ij})}\th_{ij}^{-(1+m_{ij}\a_{ij})}  
		\times \exp\left\{-\frac{b_{ij}(t^{ij}_{q_{ij}},\a_{ij})}{\th_{ij}^{\a_{ij}}}\right\}\th_{ij}^{-[1+\a_{ij}m_{ij}]}\\
	& 	\propto  \; \; \th_{ij}^{-[1+\a_{ij}(m_{ij}+N_{ij})]}
		\exp\left\{-\frac{b_{ij}(t^{ij}_{q_{ij}},\a_{ij})+\tilde{M}_{ij}(\a_{ij})}{\th_{ij}^{\a_{ij}}}\right\}.
\end{align*}
As one can see, the last function is the kernel of an inverse Gamma distribution with parameters $m_{ij}+N_{ij}$ and $b_{ij}(t^{ij}_{q_{ij}},\a_{ij})+\tilde{M}_{i,j}(\a_{ij})$ and point $b)$ follows. \newline 
A similar reasoning yields a full conditional  distribution for $\a_{ij}$ %$\pi_{3,ij}(\a_{ij}|\bj,  \bx,  u_T, \ j_{\t+1}, \bp, \ba_{(-ij)},\bth,	\bt)$ which is 
proportional to  \eqref{eq:pi3_posterior}.  
Furthermore,  concerning its log-concavity,  notice that the function in \eqref{eq:pi3_posterior} can be written as the product of the following four log-concave functions: 
\begin{align*}
	&  \a_{ij}^{N_{ij}+m_{ij}}  ,  \  \Bigl(1-\frac{\a_{ij}}{\a_{0,ij}}\Bigr)^{c_{ij}-1}  ,  \ 
		\exp\Bigl\{ - \bigl(m_{ij}d_{ij} -\log \frac{\C_{ij} (t^{ij}_{q_{ij}})^{m_{ij}}}{\th_{ij}^{N_{ij} + m_{ij}}}\bigr)\a_{ij}\Bigr\}   , \  
	\exp\Bigl\{- \frac{b_{ij}(t^{ij}_{q_{ij}},\a_{ij})+\tilde{M}_{ij}(\a_{ij})  }{\th_{ij}^{\a_{ij}}}\Bigr\} 
\end{align*}
In particular, %the first factor is log-concave since  $ N_{ij}+1+m_{ij}>0$, 
the second function is log-concave for $c_{ij} \ge 1$  and 
the last term  is a product of log-concave functions of the kind $\a_{ij} \mapsto \exp\{-z^{\a_{ij}}\}$.
Hence the log-concavity follows from the property that the product of log-concave functions is log-concave too. 

Regarding point $d)$, it is enough to observe that Equations~\eqref{eq:MRP_n}  and  \eqref{eq:full-likelihood-param} imply the following:
 \begin{align*}
P(J_{\t+1}=j |\ \bj, \bx, u_T, \bp , \ba, \bth )  & =
\frac{P(J_{\t+1}=j,  \  X_{\tau +1} >u_T  | \ \bj, \bx,\bp , \ba, \bth)}{\sum_{k\in E}P(J_{\t+1}=k,  \  X_{\tau +1} >u_T   | \ \bj, \bx,\bp , \ba, \bth)} \\
& =\frac{p_{j_{\t}j}\exp\left\{-\left(\frac{u_T}{\th_{j_{\t}j}}\right)^{\a_{j_{\t}j}}\right\}}{\sum_{k\in E}
p_{j_{\t}k}\exp\left\{-\left(\frac{u_T}{\th_{j_{\t}k}}\right)^{\a_{j_{\t}k}}\right\}}  \ . 
\end{align*}
Finally, if $t^{ij}_{q_{ij}}$ is  uniformly distributed  over $(t_1, t_2)$, then
\begin{align*}
\pi(t_{q_{ij}}^{ij}  &|  \ \bj, \bx,    u_T,   \bp,  \ba,  \bth, \bt_{{\bq}(-ij)})  \propto  
\pi_2(\th_{ij}|\a_{ij},t^{ij}_{q_{ij}})\indic(t_1 \le t_{q_{ij}}^{ij} \le t_2)	\\
&\propto b^{m_{ij}}(t^{ij}_{q_{ij}},\a_{ij})\exp\left\{-\frac{b(t^{ij}_{q_{ij}},\a_{ij})}{\th_{ij}^{\a_{ij}}}\right\}\indic(t_1 \le t_{q_{ij}}^{ij} \le t_2) \\
&\propto \left(t_{q_{ij}}^{ij}\right)^{\a_{ij}m_{ij}} \exp\left\{-\left(\frac{t_{q_{ij}}^{ij}}{\th_{ij}[(1-q_{ij})^{-1/m_{ij}}-1]^{1 /\a_{ij}}}\right)^{\a_{ij}} \right\} 
			\indic(t_1 \le t_{q_{ij}}^{ij} \le t_2)
	\end{align*}
The last equation is the kernel of a generalized Gamma, doubly-truncated at $(t_1, t_2)$, as introduced in \cite{Stacy}, so point $e)$ follows.
\end{proof}

\subsection{JAGS implementation} Proposition~\ref{prop:A.1} implies that JAGS should be able to run an exact Gibbs sampler.
The model description we have adopted in the JAGS language is based on the full likelihood \eqref{eq:full-likelihood-param}.
It is not important that the model description matches the actual model which generated the data as long as the full conditional distributions, which are determined by the joint distribution of the data and the parameters, remain unchanged. In detail, we consider the following joint distribution:
$$
L_1(\bj,\bx,u_T|j_{\tau+1},\bp,\ba,\bth) \pi(j_{\tau+1}|\bp_{j_\tau}) \pi(\bp,\ba,\bth)
$$
where $\pi(\bp,\ba,\bth)$ is the joint prior as derived in Section~\ref{sec:3} using Equations \eqref{eq:dirich}, \eqref{pi-2} and \eqref{pi-3} and 
\begin{align*}
\pi(j_{\tau+1}|\bp_{j_\tau}) & = p_{j_\tau j_{\tau+1}}\\
L_1(\bj,\bx,u_T|j_{\tau+1},\bp,\ba,\bth) & = L(\bj,\bx,u_T,j_{\tau+1}|\bp,\ba,\bth)/p_{j_\tau j_{\tau+1}}
\end{align*}

The factors of the likelihood $L_1$, to be extracted from Equation \eqref{eq:full-likelihood-param}, are modelled in JAGS as follows.
For any value of $i$, the factor
$$
\prod_{k\in E} p_{ik}^{N_{ik}}
$$
is contributed by a multinomial likelihood with probability vector $\bp_i$ and $\sum_k N_{ik}$ trials.
The factors in square brackets in \eqref{eq:full-likelihood-param} are contributed by the uncensored Weibull holding times for every string $(i,k)$ and are obtained in JAGS as Weibull densities with parameters $\alpha_{ik}$ and $\theta_{ik}$, using a ``for'' loop sweeping the strings. The last factor, which accounts for the censored holding time $u_T$, is handled by a special instruction, by which $u_T$ is first declared to be a right censored time with upper censoring point $T-t_\tau$, and then is assigned a Weibull distribution with parameters $\alpha_{j_\tau j_{\tau+1}}$ and $\theta_{j_\tau j_{\tau+1}}$.

The factor $\pi(j_{\tau+1}|\bp_{j_\tau}) \pi(\bp,\ba,\bth)$, representing the prior associated with $L_1$, is handled as follows. The additional prior on $j_{\tau+1}$ is a discrete distribution on the integers 1, 2, 3, with probabilities taken from the row of $\bp$ indexed by $j_\tau$. Every row $\bp_i$ of $\bp$ is assigned a Dirichled distribution directly. Whenever $m_{ik}\ge 2$, as $c_{ik}=m_{ik}$, the shape parameters $\alpha_{ik}$ have a shifted Gamma prior, see Equation \eqref{pi-3}, obtainable by defining in JAGS a new non-shifted Gamma variable with shape $c_{ik}$ and rate $m_{ik}d_{ik}$, which, after summing the shift, is assigned to $\alpha_{ik}$; for the value of $d_{ik}$ see Equation \eqref{eq:d_ij}. The generalized Gamma for $\theta_{ik}$ is defined conditionally on $\alpha_{ik}$: first a Gamma prior with shape $m_{ij}$ and rate $\hat{b}_{ik}(\alpha_{ik})$ is assigned to a new random variable $a_{ik}$ and then $a_{ik}^{-1/\a_{ik}}$ is assigned to $\theta_{ik}$. 

In case there is either just one observation or  no learning dataset $\by_m$, some special instructions in the JAGS code are needed. 
In particular, if there is no  learning dataset, then the missing learning dataset is substituted, for every string $(i,k)$, by the fictitious observation $\tilde{t}_{ik}$ drawn from a uniform distribution over $(1,5000)$ days (so $m_{ik}=1$ for all strings). Then, the priors of the $\theta_{ik}$ retain the same form, whereas the priors of the $\alpha_{ik}$'s, derived from Equation \eqref{pi-3} with $m_{ik}=1$ and $c=2$ (a value taken from Table~\ref{table:hyperparameters}), are
$$
\pi_3(\alpha_{ik}) \propto \left(1-\frac{\alpha_{0,ik}}{\alpha_{ik}}\right) \indic(\alpha_{0,ik}\le\alpha_{ik}\le\alpha_{1,ik})
$$
with assigned $\alpha_{0,ik}$ and $\alpha_{1,ik}$.
This latter distribution is coded using the so-called zeros trick: a fictitious zero observation from a Poisson distribution with mean $\phi_{ik}=-\ln \pi_3(\alpha_{ik})$ is introduced; then a uniform prior over $[\a_{0,ik},\a_{1,ik}]$ is assigned to $\alpha_{ik}$. The effect on the formula of the joint distribution is that the likelihood $L_1$ gets multiplied by the factor $\exp(-\phi_{ik})$, contributed by the zero observation; the multiplication by the uniform density gives back the correct factor accounting for the prior of $\alpha_{ik}$.

\section*{\large{Acknowledgments}}
We are grateful to Renata Rotondi for providing us with data and the map of Italy along with her very helpful comments. We are solely responsible for any remaining inaccuracy.

\end{document}